\documentclass{cernrep}

\usepackage{xcolor}

\usepackage{fancyhdr}
\usepackage[colorlinks=true, linkcolor=black, citecolor=black, filecolor=black, urlcolor=blue]{hyperref} 
\fancyhfoffset{4 mm}
\fancypagestyle{ARTTITLE}{%
\fancyhf{} 
\lhead{\small{Proceedings of the 2018 CERN--Accelerator--School course on\\
\it{Numerical Methods for Analysis, Design and Modelling of Particle Accelerators}, Thessaloniki, (Greece)}} 
\lfoot{Available online at \url{https://cas.web.cern.ch/previous-schools}}
\rfoot{\thepage\hspace*{3mm}}
 
}

\begin{document}
  
  \title{Magnetodynamic Finite-Element Simulation of Accelerator Magnets}
  \author{H. De~Gersem, I. Cortes Garcia, L.A.M. D'Angelo and S. Sch\"ops\thanks
    {This lectures partially originates in the collaboration project \emph{Simulation of Transient Effects in Accelerator Magnets} (STEAM, https://espace.cern.ch/steam) between CERN and TEMF. This work has been supported by the Excellence Initiative of the German Federal and State Governments and the Graduate School of Computational Engineering at TU Darmstadt}}
  \institute{Institute for Accelerator Science and Electromagnetic Fields (TEMF), TU Darmstadt, Germany}

  \begin{abstract}
This lecture note describes how to set up and what is behind a magnetodynamic field simulation for an accelerator magnet. The relevant formulation of Maxwell's equations is derived. The formulation is discretized in space by the finite-element method and in time by a standard time integration method. The~steps for setting up the accelerator-magnet model are described. An exemplary simulation of the GSI SIS-100 magnet is given as illustration. Finally, some extensions to the standard FE method, dedicated to accelerator magnets, are discussed.
  \end{abstract}
  
  \keywords{Accelerator magnets; finite-element method; electromagnetic field simulation.}
  
  \maketitle
  \thispagestyle{ARTTITLE}
  
\section{Introduction}

Contemporary accelerator magnets are designed with the help of finite-element (FE) field simulation. To that purpose, several software packages such as, e.g., Opera \cite{Opera_2019aa}, ANSYS Maxwell \cite{Maxwell_2019aa}, MagNet \cite{MagNet_2019aa}, Flux \cite{AltairFlux_2019aa} and CST EM STUDIO\textregistered \cite{CST_2017aa}, are commercially available. Several accelerator laboratories dispose of an own tool, such as, e.g. ROXIE at CERN \cite{ROXIE_2019aa,Russenschuck_2010aa}. Despite the well-performing solvers and the highly intuitive graphical user interfaces, modelling and simulating accelerator magnets remains a tedious task. This lecture note addresses physicists and engineers which are new in the domain of magnetic field simulation or in the discipline of accelerator science. The note is deliberately held simple. Nevertheless, digging further into theory or practice is possible by following the references.

In Section~\ref{sect:hdg_mqs}, the magnetoquasistatic formulation is derived from the full set of Maxwell equations. Section~\ref{sect:hdg_space} is devoted to discretizing the field equation in space by the finite-element (FE) method. Section~\ref{sect:hdg_time} gives is short notice on time integration. Solving the resulting algebraic system of equations is the computionally most expensive part of a FE solver and is addressed in the very short Section~\ref{sect:hdg_system_solution}. From then on, information which is more specific for accelerator-magnet simulation, is given. Section~\ref{sect:hdg_linearization} deals with modelling laminated yoke parts and resolving ferromagnetic saturation. In Section~\ref{sect:hdg_postprocessing}, typical post-processing actions needed to obtain the relevant performance parameters of an accelerator magnet are discussed. Section~\ref{sect:hdg_modelling} gives information about setting up an FE magnet model, whereas Section~\ref{sect:hdg_example} describes the simulation of the SIS-100 magnet as an example. Section~\ref{sect:hdg_advanced} shows some recent developments aiming at further improvement of 3D magnet simulation. The note ends with a short summary.

\section{Magnetoquasistatic formulation}
\label{sect:hdg_mqs}

\begin{figure}[tb]
  \centering
  (a)\includegraphics[width=4.5cm]{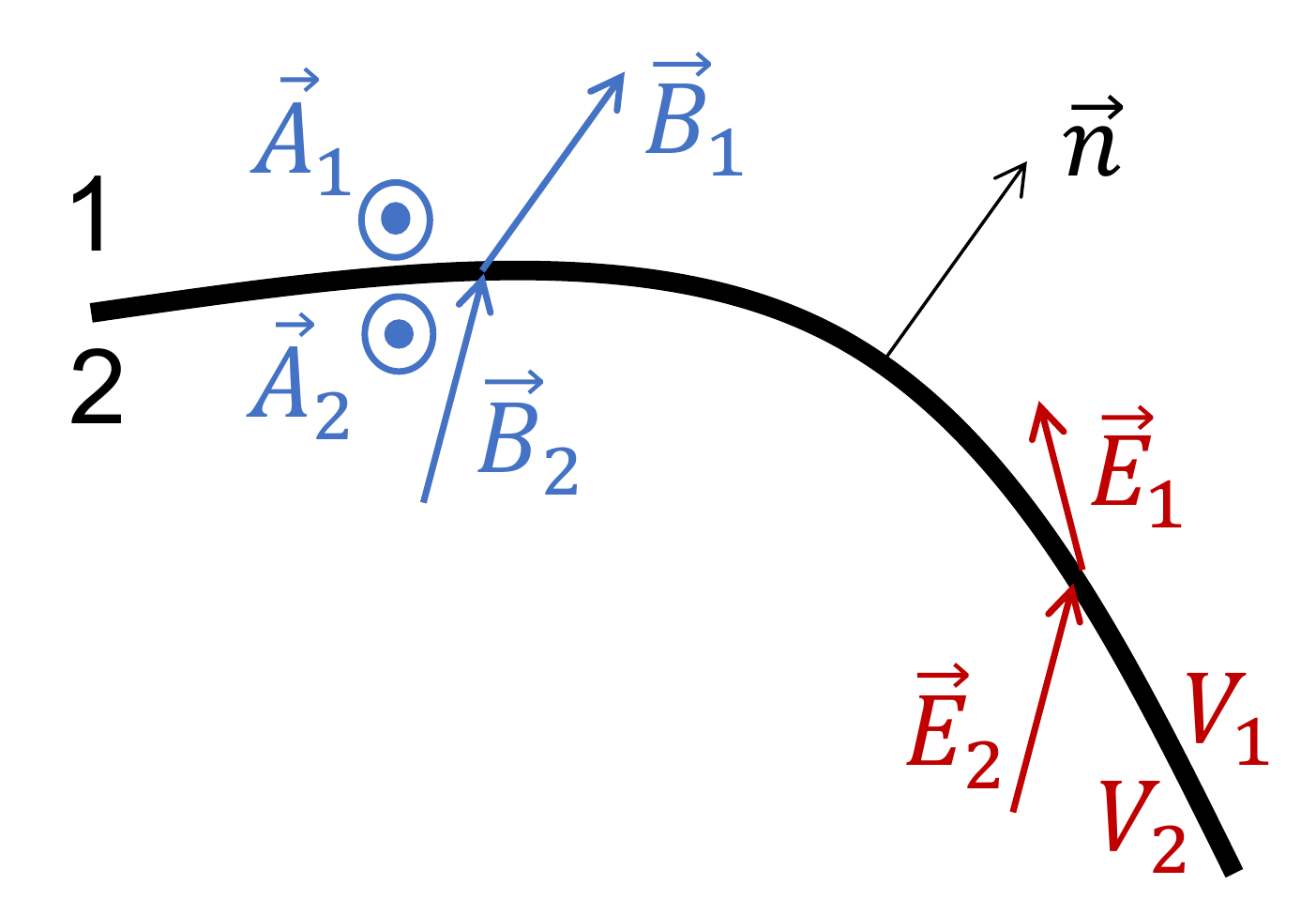}
  \hspace{1cm}(b)\includegraphics[width=4.5cm]{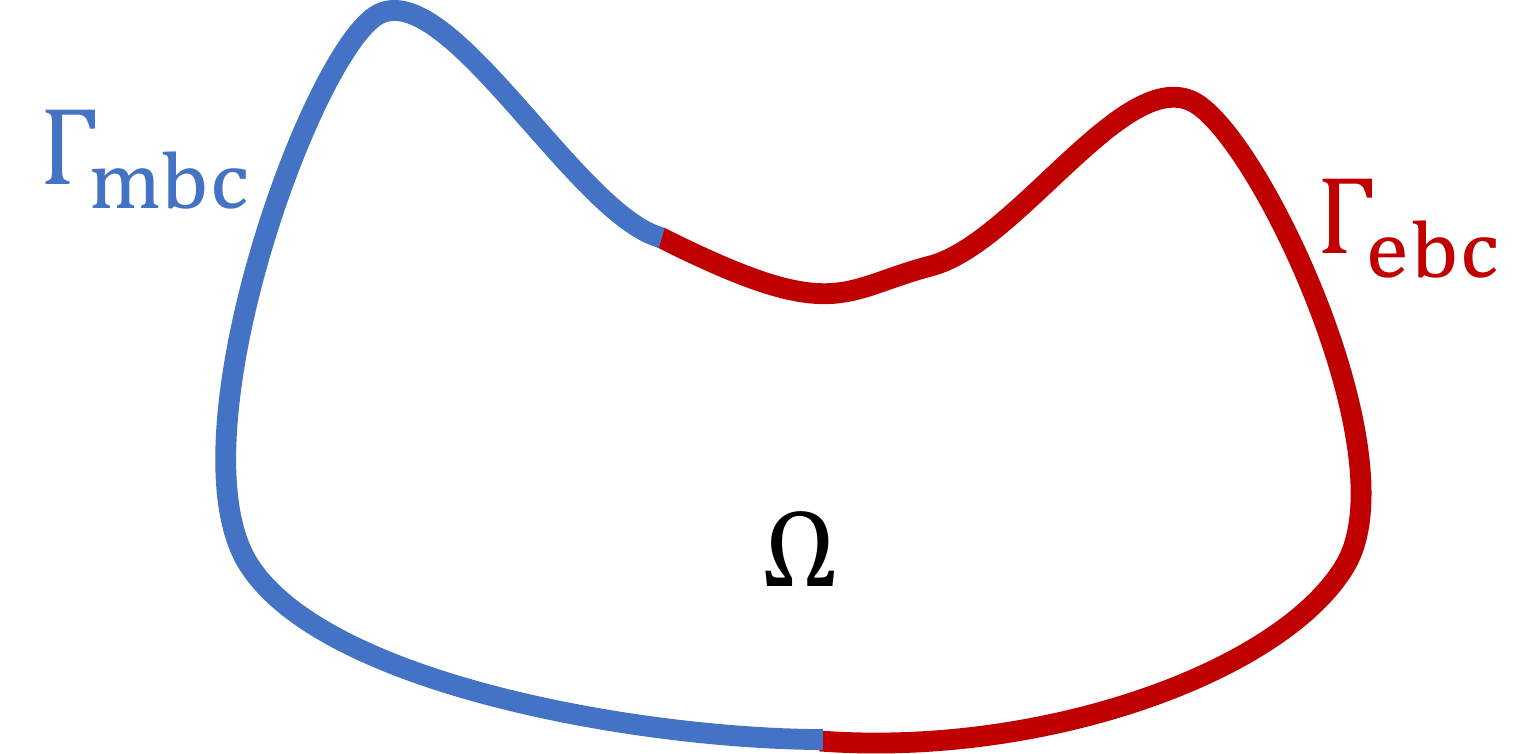}
  \caption{(a) Interface between region~1 and region~2 with normal vector $\vec{n}$; (b) computational domain $\Omega$, its boundary $\partial\Omega=\Gamma_\text{ebc}\cup\Gamma_\text{mbc}$ with disjunct parts $\Gamma_\text{ebc}$ and $\Gamma_\text{mbc}$ at which electric and magnetic boundary conditions are applied, respectively.}
  \label{fig:hdg_interface}
\end{figure}

Accelerator magnets are excited by currents that vary slowly in time, which allows to neglect displacements currents with respect to conducting and magnetic effects \cite{Haus_1989aa,Dirks_1996aa}. Hence, the relevant subset of Maxwell's equations is
\begin{subequations}
  \begin{alignat}{3}
    \label{eq:hdg_magngauss} \nabla\cdot\vec{B} &=0
    &&\quad\Leftarrow\quad & \vec{B} &= \nabla\times\vec{A} \,;\\
    \label{eq:hdg_faraday} \nabla\times\vec{E} &= -\frac{\partial\vec{B}}{\partial t}
    &&\quad\Leftarrow\quad & \vec{E} &=-\frac{\partial\vec{A}}{\partial t}-\nabla V \,;\\
    \label{eq:hdg_ampere} \nabla\times\vec{H} &= \vec{J}\,, && &
  \end{alignat}
\end{subequations}
where $\vec{B}(\vec{r},t)$ is the magnetic flux density, $\vec{E}(\vec{r},t)$ is the electric field strength, $\vec{H}(\vec{r},t)$ is the magnetic field strength, $\vec{J}(\vec{r},t)$ is the current density, $\vec{r}$ is the spatial coordinate and $t$ is the time \cite{Maxwell_1864aa,Jackson_1998aa,Griffiths_1999aa}. The~magnetic Gauss law \eqref{eq:hdg_magngauss} is enforced by the definition of the magnetic vector potential $\vec{A}(\vec{r},t)$. Also Faraday's law \eqref{eq:hdg_faraday} is integrated in space, which leads to the introduction of the electric scalar potential $V(\vec{r},t)$. The~Maxwell laws comes together with particular interface conditions, i.e.,
\begin{subequations}
  \begin{alignat}{3}
  \label{eq:hdg_magngauss2} \vec{n}\cdot\vec{B}_1 &=\vec{n}\cdot\vec{B}_2
  &&\quad\Leftarrow\quad & \vec{n}\times\vec{A}_1 &= \vec{n}\times\vec{A}_2 \,;\\
  \label{eq:hdg_faraday2} \vec{n}\times\vec{E}_1 &=\vec{n}\times\vec{E}_2
  &&\quad\Leftarrow\quad & V_1 &=V_2+\text{ct} \,,
  \end{alignat}
\end{subequations}
where $\vec{n}$ is a unit vector normal to the interface between two regions indicated by subscripts~$1$ and $2$ (Fig.~\ref{fig:hdg_interface}). The interface conditions carry over in a particular way to the defined potentials, see Eqs. \eqref{eq:hdg_magngauss2} and \eqref{eq:hdg_faraday2}.

The behaviour of the present materials and excitations is described by the constitutive equations
\begin{subequations}
\begin{align}
  \label{eq:hdg_sigma} \vec{J} &=\sum_{q=1}^{n_\text{coil}} \vec{\chi}_q i_q +\sigma\vec{E} \,;\\
  \label{eq:hdg_nu} \vec{H} &=\vec{H}_\text{c} +\nu\vec{B} \,,
\end{align}
\end{subequations}
where $\vec{\chi}_q(\vec{r})$ is a winding function modelling the spatial distribution of the currents $i_q(t)$ in each of the~$n_\text{coil}$ coils of the magnet, $\sigma(\vec{r})$ is the conductivity, $\vec{H}_\text{c}(\vec{r},t)$ is the coercivity of hard or soft magnetic material, $\nu(\vec{r})=\mu^{-1}(\vec{r})$ is the reluctivity and $\mu(\vec{r})$ is the permeability. For nonlinear materials, the reluctivity, permeability and coercitivity depend on the magnetic field, i.e., $\nu=\nu(\vec{r},\vec{B}(\vec{r}))$, $\mu=\mu(\vec{r},\vec{H}(\vec{r}))$ and $\vec{H}_\text{c}(\vec{r},\vec{B}(\vec{r}))$.

The magnetoquasistatic formulation in terms of the magnetic vector potential and the electric scalar potential, also called the $\vec{A}$-$V$ formulation \cite{Kameari_1990aa,Biro_1989aa,Bossavit_1998aa}, is found by combining Eqs. \eqref{eq:hdg_ampere}, \eqref{eq:hdg_sigma} and \eqref{eq:hdg_nu}:
\begin{subequations}
\begin{align}
  \label{eq:hdg_Avform1}\nabla\times\left(\nu\nabla\times\vec{A}\right)
  +\sigma\frac{\partial\vec{A}}{\partial t} +\sigma\nabla V
  &=\sum_{q=1}^{n_\text{coil}}\vec{\chi}_q i_q -\nabla\times\vec{H}_\text{c} \,;\\
  \label{eq:hdg_Avform2}-\nabla\cdot\left(\sigma\frac{\partial\vec{A}}{\partial t}\right)
  -\nabla\cdot\left(\sigma\nabla V\right) &=0 \,,
\end{align}
\end{subequations}
where the second equation is the continuity equation which also can be found by taking the divergence of the first equation.

When $(\vec{A},V)$ denotes a solution of Eqs. \eqref{eq:hdg_Avform1} and \eqref{eq:hdg_Avform2}, $(\vec{A}+\nabla\psi,V-\frac{\partial\psi}{\partial t})$ is a solution of the~formulation as well. This indicates a lack of uniqueness, which can be alleviated by applying a~so-called \emph{gauge condition} \cite{Biro_1989aa,Clemens_2002aa,Clemens_2011aa}. Numerous possibilities exist. One of the most common gauged formulations is the $\vec{A}^*$-formulation, where $V$ is chosen to be zero and the divergence of $\vec{A}$ is fixed in the~non-conducting model part $\Omega_\text{nc}=\{\vec{r}:\sigma(\vec{r})=0\}$, e.g., by the Coulomb gauge $\nabla\cdot\vec{A}(\vec{r})=0$, $\vec{r}\in\Omega_\text{nc}$. The resulting partial differential equation (PDE) is then
\begin{align}
  \label{eq:hdg_Astarform}\nabla\times\left(\nu\nabla\times\vec{A}\right)
  +\sigma\frac{\partial\vec{A}}{\partial t}
  =\sum_{q=1}^{n_\text{coil}}\vec{\chi}_q i_q -\nabla\times\vec{H}_\text{c} \,,
\end{align}
which will be the formulation used further on. Equation~(\ref{eq:hdg_Astarform}) is a parabolic PDE and reduces to an elliptic PDE for the case where $\sigma=0$. The formulation is considered at a computational domain~$\Omega$ with boundary $\partial\Omega=\Gamma_\text{ebc}\cup\Gamma_\text{mbc}$ consisting of two disjunct parts: $\Gamma_\text{ebc}$ at which the homogeneous \emph{electric boundary condition} $\vec{n}\times\vec{A}=0$ and $\Gamma_\text{mbc}$ at which the homogeneous \emph{magnetic boundary condition} $\vec{n}\times\vec{H}=0$ is applied (Fig.~\ref{fig:hdg_interface}b).

Many other formulations using variants of the magnetic vector potential and the electric scalar potential exist. Moreover, a complete other set of formulations, known as $\vec{T}-\Omega$-formulations, using an electric vector potential and a magnetic scalar potential \cite{Webb_1993aa,Biro_1995aa} exist as well and can be applied for accelerator-magnet simulation. Further information on formulations can be found in \cite{Bossavit_1998aa}.

\section{Discretization in space}
\label{sect:hdg_space}

\subsection{Ritz-Galerkin weighted-residual approach}
\label{subsect:hdg_ritz_galerkin}

The presence of heterogeneous and nonlinear materials necessitates the use of a volumetric spatial discretization technique as, e.g., the finite-element (FE) method \cite{Silvester_1996aa,Brenner_2008aa}, the finite-difference method \cite{Kunz_1993aa}, the finite integration technique (FIT) \cite{Weiland_1977aa,Weiland_1996aa,Clemens_2005aa} or a spectral-element method \cite{Boyd_2001aa}. The FE method is the~most popular one and will be developed here. The key point is to replace the PDE \eqref{eq:hdg_Astarform}, which should apply to each point in space, by its weak form, i.e., integrated according to $N$ test functions $\color{red}\vec{w}_i(\vec{r})$, $i=1,\ldots,N$ (which will be specified below) on the computational domain~$\Omega$. The weak form is
\begin{align}
\nonumber\lefteqn{\int_\Omega \nabla\times\left(\nu\nabla\times\vec{A}\right)\cdot\textcolor{red}{\vec{w}_i}\,\mbox{d}V
+\int_\Omega\sigma\frac{\partial \vec{A}}{\partial t}\cdot\textcolor{red}{\vec{w}_i}\,\mbox{d}V} & \\
  &=\sum_{q=1}^{n_\text{coil}}i_q \int_\Omega\vec{\chi}_q\cdot\textcolor{red}{\vec{w}_i}\,\mbox{d}V
-\int_\Omega\nabla\times\vec{H}_\text{c}\cdot\textcolor{red}{\vec{w}_i}\,\mbox{d}V \,.
\end{align}
A bit of vector calculus brings up
\begin{align}
\nonumber\lefteqn{\int_{\partial\Omega}\left((\nu\nabla\times\vec{A})\times\textcolor{red}{\vec{w}_i}\right)\cdot\,\mbox{d}\vec{S}
  +\int_\Omega\left(\nu\nabla\times\vec{A}\right)\cdot\left(\nabla\times\textcolor{red}{\vec{w}_i}\right)\,\mbox{d}V +\int_\Omega\sigma\frac{\partial \vec{A}}{\partial t}\cdot\textcolor{red}{\vec{w}_i}\,\mbox{d}V} & \\
\label{eq:hdg_weakform}&=\sum_{q=1}^{n_\text{coil}} i_q \int_\Omega\vec{\chi}_q\cdot\textcolor{red}{\vec{w}_i}\,\mbox{d}V
-\int_{\partial\Omega}\left(\vec{H}_\text{c}\times\textcolor{red}{\vec{w}_i}\right)\cdot\,\mbox{d}\vec{S}
+\int_\Omega\vec{H}_\text{c}\cdot\nabla\times\textcolor{red}{\vec{w}_i}\,\mbox{d}V
 \,.
\end{align}
When only homogeneous electric boundary conditions (BCs) and homogeneous magnetic BCs are present and when no permanent magnets are positioned alongside the boundary ($\vec{H}_\text{c}=0$ on $\partial\Omega$), both boundary integral terms vanish. For other configurations, we refer to literature, e.g., \cite{Bastos_2003aa}.

The FE procedure continues with discretizing the magnetic vector potential, i.e., expressing it as a~linear combination of $N$ \emph{trial functions} $\textcolor{blue}{\vec{w}_j(\vec{r})}$, $j=1,\ldots,N$, here chosen identically to the test functions (Ritz-Galerkin approach),
\begin{align}
  \label{eq:hdg_Adisc} \vec{A}\approx \vec{A}_h(\vec{r},t) &=\sum_{j=1}^N u_j(t)\textcolor{blue}{\vec{w}_j(\vec{r})} \,,
\end{align}
where $u_j(t)$ are the degrees of freedom (DoFs) before time discretization. The subscript $h$ distinguishes between the exact solution and the discrete solution but will be omitted in all further development for reasons of conciseness. Combined with Eq. \eqref{eq:hdg_weakform}, we find the system of equations
\begin{align}\label{eq:hdg_mqssystem}
  K_\nu u +M_\sigma\frac{\mbox{d}u}{\mbox{d}t} &=X i +g \,,
\end{align}
where the coefficients of the contributing matrices are calculated from
\begin{subequations}
  \begin{align}
  \label{eq:hdg_Knuij}K_{\nu,ij} &=\int_\Omega \left(\nu\nabla\times\textcolor{blue}{\vec{w}_j}\right) \cdot\left(\nabla\times\textcolor{red}{\vec{w}_i}\right) \,\mbox{d}V \,;\\
  \label{eq:hdg_Msigmaij}M_{\sigma,ij} &=\int_\Omega \sigma\textcolor{blue}{\vec{w}_j}\cdot\textcolor{red}{\vec{w}_i}\,\mbox{d}V \,;\\
  \label{eq:hdg_Xik}X_{ik} &=\int_\Omega \vec{\chi}_k\cdot\textcolor{red}{\vec{w}_i}\,\mbox{d}V \,;\\
  \label{eq:hdg_Hci}g_{ik} &=\int_\Omega \vec{H}_\text{c}\cdot\nabla\times\textcolor{red}{\vec{w}_i}\,\mbox{d}V \,.
  \end{align}
\end{subequations}
Note that $\Omega_\text{nc}$ may be not empty and thus the matrix $M_\sigma$ in front of the time derivative is not invertible. System like that are called \emph{differential-algebraic} because some equations ''lack a derivative'' \cite{Griepentrog_1986aa}.

\subsection{Edge FE shape functions}

\begin{figure}[tb]
  \centering
  (a)\includegraphics[width=4.5cm]{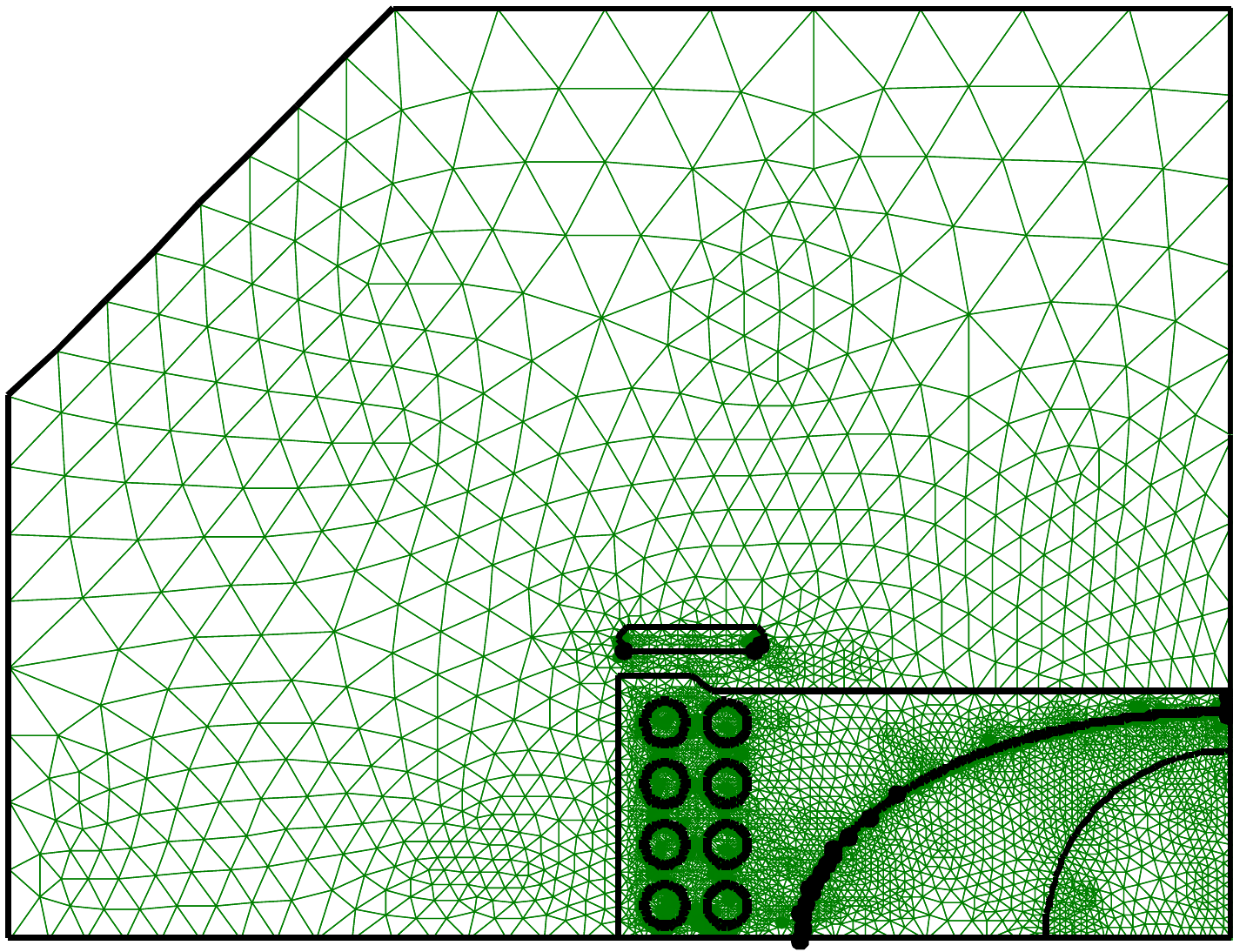}
  \hspace{1cm}(b)\includegraphics[width=8.5cm]{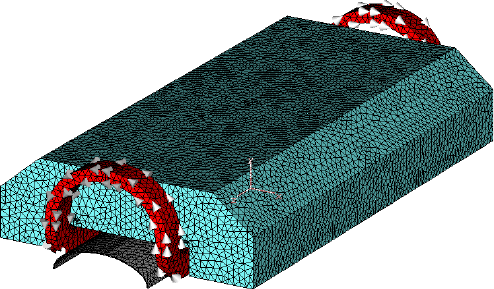}
  \caption{(a) 2D mesh and (b) 3D mesh of the SIS-100 magnet model.}
  \label{fig:sis100_mesh}
\end{figure}

The FE method is typically employed on a tetrahedral mesh in the 3D case or on a 2D triangular mesh in the 2D case (Fig.~\ref{fig:sis100_mesh}). Each tetrahedron/triangle is called a \emph{mesh element}. It is assumed that the mesh resolves the material distribution, i.e., each element contains a single material. The test and trial functions needed in Section~\ref{subsect:hdg_ritz_galerkin} are defined element-wise.

\begin{figure}[tb]
  \centering
  (a)\includegraphics[width=4cm]{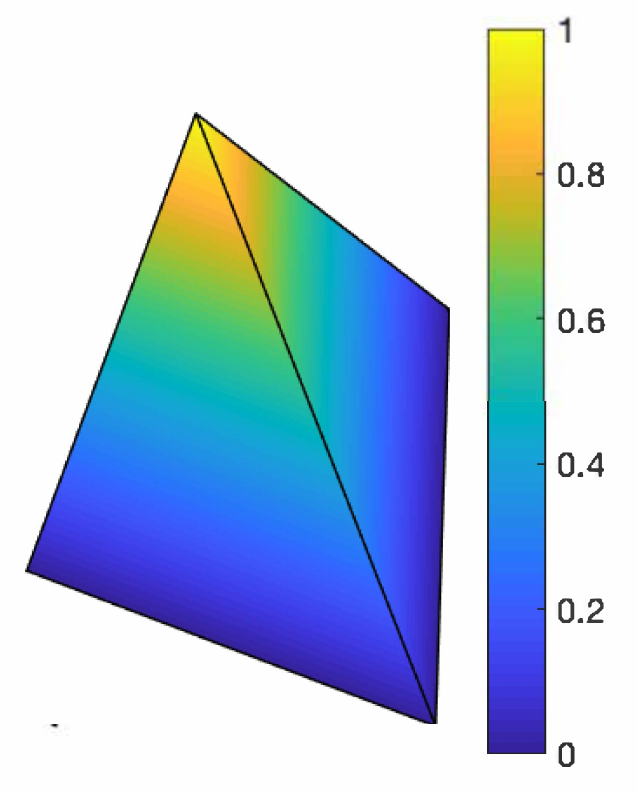}
  (b)\includegraphics[width=5cm]{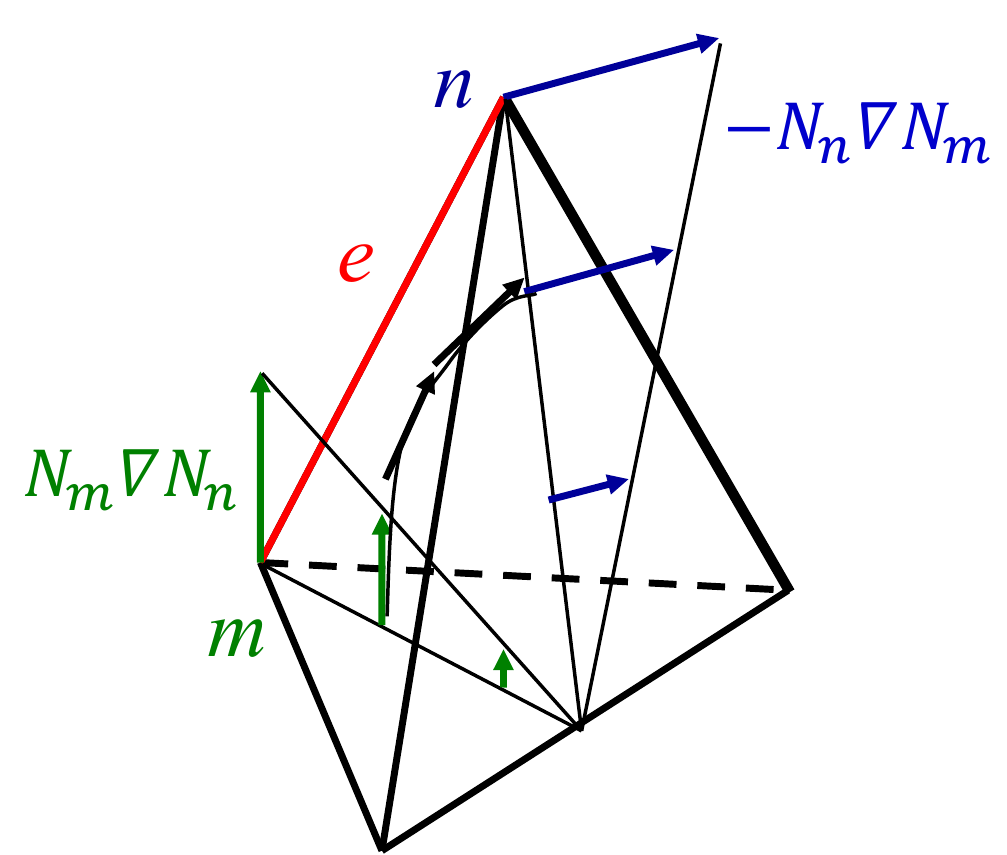}
  \caption{(a) Nodal FE shape function $N_n(\vec{r})$ and (b) edge FE shape function $\vec{w}_i(\vec{r})$ constructed as the sum of $\textcolor{green}{N_m\nabla N_n}$ and $\textcolor{blue}{-N_n\nabla N_m}$.}
  \label{fig:hdg_FEshapefunctions}
\end{figure}

According to Eq. \eqref{eq:hdg_magngauss2}, the magnetic vector potential $\vec{A}(\vec{r},t)$ should be tangentially continuous at material interfaces but the normal components may jump. Because neighbouring element may feature different materials, $\vec{A}(\vec{r},t)$ should be tangentially continuous at the interfaces between the elements. A~convenient strategy is to define so-called \emph{edge functions}, which are vectorial functions associated with the edges, enforcing tangential continuity (Fig.~\ref{fig:hdg_FEshapefunctions}(b)). By construction, the tangential continuity carries over to the faces between the elements. There exist canonical procedures to set up such functions, which typically start from a set of scalar functions, called \emph{nodal functions}, which are continuous at the element boundaries.


\subsubsection{3D case on a tetrahedral mesh}

In the 3D case, lowest-order edge functions for tetrahedra are constructed as follows. Piecewise linear nodal functions $N_m(\vec{r})$ are defined which take the value~$1$ at node~$m$ and linearly decay to~$0$ along each of the edges incident to node~$m$ (Fig.~\ref{fig:hdg_FEshapefunctions}(a)). Such functions can be defined element-wise on a tetrahedron~$\Omega_k\subset\Omega$ incident to node~$m$ by the expression
\begin{align}
  N_m(x,y,z) &=a_{k,m}+b_{k,m}x+c_{k,m}y+d_{k,m}z \,,
  \quad\text{for }(x,y,z)\in\Omega_k \,,
\end{align}
where the coefficients $a_{k,m}$, $b_{k,m}$, $c_{k,m}$, $d_{k,m}$ are determined such that $N_m(\vec{r}_n)=\delta_{mn}$ for any two corner nodes~$m$ and $n$ of tetrahedron~$k$.

\begin{figure}[tb]
  \centering
  \includegraphics[width=8cm]{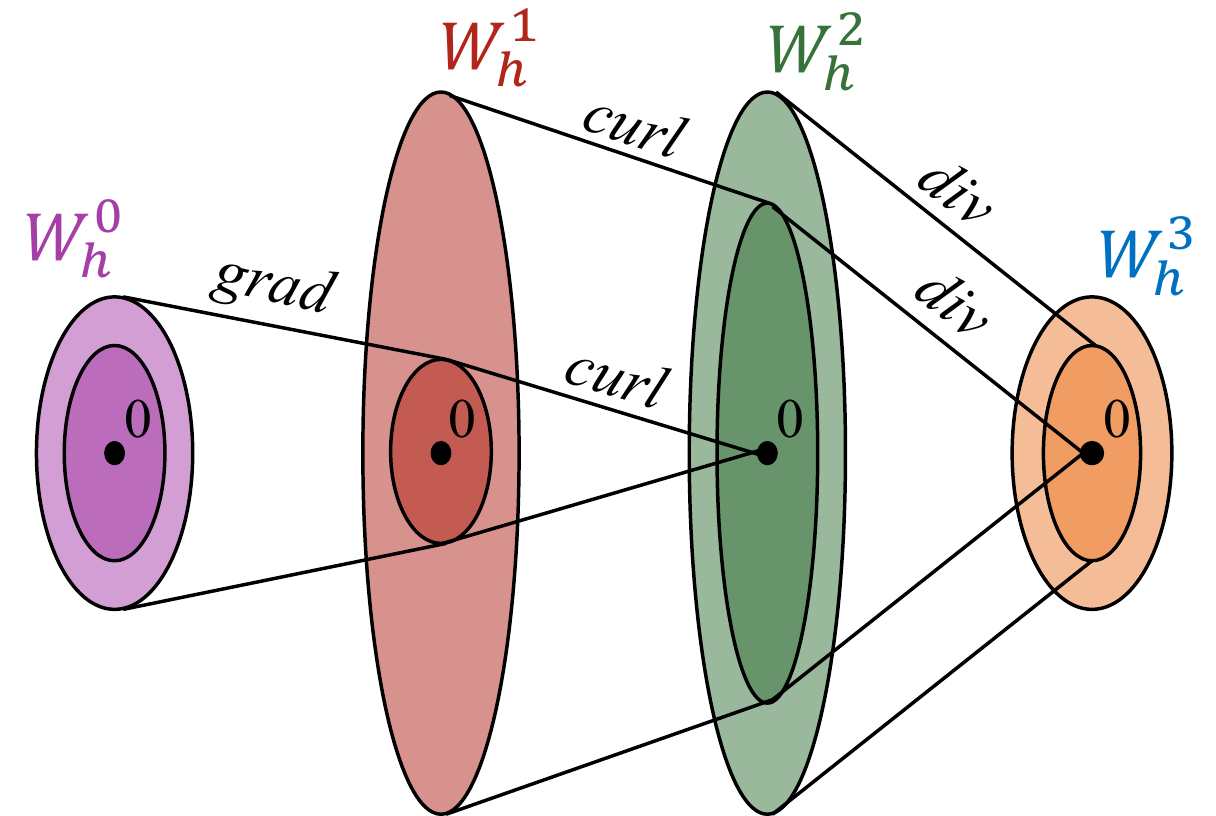}
  \caption{Whitney complex: $W_h^0$ denotes the set of discrete nodal shape functions, $W_h^1$ the set of discrete edge shape functions, $W_h^2$ the set of discrete face shape functions and $W_h^3$ the set of discrete volume shape functions.}
  \label{fig:hdg_Whitneycomplex}
\end{figure}

The corresponding set of edge functions $\vec{w}_i$ are defined by (Fig.~\ref{fig:hdg_FEshapefunctions}b)
\begin{align}
  \label{eq:hdg_edgeshapefunction}
  \vec{w}_i(\vec{r}) &=\textcolor{green}{N_m\nabla N_n}\textcolor{blue}{-N_n\nabla N_m} \,,
  \quad\text{for }\vec{r}\in\Omega_k \,.
\end{align}
By construction, the edge functions are tangentially continuous at the element interfaces. Moreover, they fulfil a \emph{partition-of-unity} property formalized by
\begin{align}
  \int_{L_j} \vec{w}_i\cdot\,\mbox{d}\vec{s} &=\delta_{ij} \,,
\end{align}
where $L_j$ denotes the edge~$j$ of the mesh. Furthermore, the nodal and edge function spaces form a part of a discrete Whitney complex (Fig.~\ref{fig:hdg_Whitneycomplex}). The finite set $W_h^0\subset W^0$ of nodal functions, being a subset of the set of continuous scalar functions $W^0$, is mapped by the gradient operator onto a subset of the set $W_h^1$ of edge functions ($\text{grad}\, \psi_h\in W_h^1$, $\forall\psi_h\in W_h^0$), which is on its turns mapped by the curl operator onto $0$ ($\text{curl}\, \text{grad}\, \psi_h=0$, $\forall\psi_h\in W_h^0$). This mimics the property $\nabla\times\nabla\psi=0$, $\forall\psi\in W^0$ at the discrete level. For further information, the reader is referred to, e.g., \cite{Bossavit_1998aa}.

\subsubsection{2D case on a triangular mesh}

Many magnet systems allow the use of a 2D FE simulation set-up, at least during the first stages of the design. One distinguishes between the Cartesian case, where the cross-sectional $(x,y)$-plane of the~magnet remains invariant over a length $\ell_z$ along the $z$-direction, and the axisymmetric case where the~cross-sectional $(r,z)$-plane preserves its form under rotation by $2\pi$ along the azimuthal $\varphi$-direction. In both cases, the magnetic flux lines lie within the cross-sectional plane, from which is can be concluded that the magnetic flux density $\vec{B}=\nabla\times\vec{A}$ can be derived from a magnetic vector potential with a single component: $\vec{A}=(0,0,A_z)$ or $\vec{A}=(0,A_\varphi,0)$ for the Cartesian or axisymmetric case, respectively. It~is recommended to stick to the 3D weak formulation Eq. \eqref{eq:hdg_weakform} and to code the symmetry into the choice of FE shape functions, instead of reducing the 3D PDE to a 2D PDE which is then discretized anew. In~any case, only a 2D triangulation needs to be constructed and the calculation of the matrix coefficients \eqref{eq:hdg_Knuij}-\eqref{eq:hdg_Xik} can be carried out on the 2D mesh, which yields a considerable reduction of computational effort.

Appropriate lowest-order FE edge shape functions are \cite{Vanoost_2013aa}
\begin{subequations}
  \begin{align}
    \label{eq:hdg_wi_cart}\textcolor{red}{\vec{w}_i^\text{(cart)}}
    &=\frac{\tilde{a}_{k,i}+\tilde{b}_{k,i}x  +\tilde{c}_{k,i}y}{\ell_z}
    \vec{e}_z \,;\\
    \label{eq:hdg_wi_axi} \textcolor{red}{\vec{w}_i^\text{(axi)}}
    &=\frac{\tilde{a}_{k,i}+\tilde{b}_{k,i}r^2+\tilde{c}_{k,i}z}{2\pi r}
    \vec{e}_\varphi \,.
  \end{align}
\end{subequations}
Here, the edge with index~$i$ is a line or circle perpendicular to the cross-section plane in the node with index~$i$. Moreover, element~$k$ can be seen as a triangular prism or a triangular torus obtained by extruding the cross-sectional triangle along the $z$- or $\varphi$-direction. The nominators of Eqs. \eqref{eq:hdg_wi_cart} and \eqref{eq:hdg_wi_axi} can be interpreted as nodal shape functions defined on the cross-sectional plane taking the value~$1$ at one of the nodes and the value~$0$ at all other nodes (which fixes the coefficients $\tilde{a}_{k,i}$, $\tilde{b}_{k,i}$ and $\tilde{c}_{k,i}$). Notice the $r^2$-term in Eq. \eqref{eq:hdg_wi_axi} which is needed for obtaining a consistent FE discretization \cite{Vanoost_2013aa}. The denominators make sure that the resulting edge functions fulfil the partition-of-unity property. In the accompanying exercise \cite{De-Gersem_2019ac,De-Gersem_2019ad}, the coefficients of the system matrix and right-hand side are calculated according to Eqs. \eqref{eq:hdg_wi_cart} and \eqref{eq:hdg_Knuij}--\eqref{eq:hdg_Hci} for the 2D Cartesian case.

\subsubsection{Higher-order FE shape functions}

The accuracy of an FE simulation can be improved by refining the FE mesh (\emph{$h$-refinement}). In case of smooth solutions, increasing the polynomial order of the FE shape functions is more efficient (\emph{$p$-refinement}). The construction procedures are substantially more complicated (see, e.g., \cite{Ainsworth_2003aa,Ingelstrom_2006aa,Schoberl_2005aa,Marsic_2015aa}). Of particular interest are the so-called \emph{hierarchical} FE spaces, in which the order can be increased successively while keeping the already existing FE shape functions unchanged. When using higher-order FE shape functions, it is recommended to use curved elements in order to improve the representation of curved boundaries and material interfaces \cite{Geuzaine_2009ab}.

\section{Discretization in time}
\label{sect:hdg_time}

The semi-discrete system \eqref{eq:hdg_mqssystem} needs to be further discretized in time. Time integrators (or time-integration methods) exist in all colours and flavours (see, e.g., \cite{Hairer_2000aa} and \cite{Hairer_1996aa}). Here, only a representative family of time integrators is worked out. When the vector of DoFs $u(t)$ is linearly interpolated between two time instants $t_k$ and $t_{k+1}$, one finds for time instant $t=(1-\theta)t_k+\theta t_{k+1}$, $\theta\in[0,1]$:
\begin{subequations}
  \begin{align}
    \Delta t_{k+1} &=t_{k+1}-t_k \,;\\
    u(t) &\approx (1-\theta)u_k +\theta u_{k+1} \,;\\
    \frac{\text{d}u}{\text{d}t} &\approx \frac{u_{k+1}-u_k}{\Delta t_{k+1}} \,;\\
    i_q(t) &\approx (1-\theta)i_q(t_k) +\theta i_q(t_{k+1}) \,;\\
    g(t) &\approx (1-\theta)g(t_k) +\theta g(t_{k+1}) \,.
  \end{align}
\end{subequations}
The solution at a next time instant is then computed by solving
\begin{align}
  \nonumber\left(K_\nu+\frac{1}{\theta\Delta t_{k+1}}M_\sigma\right)u_{k+1}
  &=X\left(i_{k+1}+\frac{1-\theta}{\theta}i_k\right)
  +g_{k+1}+\frac{1-\theta}{\theta}g_k \\
  &\qquad +\left(-\frac{1-\theta}{\theta}K_\nu+\frac{1}{\theta\Delta t_{k+1}}M_\sigma\right)u_k \,.
\end{align}
For $\theta<\frac{1}{2}$, this time integrator is only conditionally stable, i.e., when the time step $\Delta t_{k+1}$ is smaller than the so-called \emph{Courant–Friedrichs–Lewy (CFL)} time step $\Delta t_\text{CFL}$, which depends on the problem type and scales unfavourably with the mesh size. Because the parabolic PDE \eqref{eq:hdg_Astarform} represents a so-called \emph{stiff} problem, $\Delta t_\text{CFL}$ would be extremely small \cite{Hairer_2000aa,Dutine_2017ac}. If there are non-conducting regions in the domain, \eqref{eq:hdg_mqssystem} is a differential-algebraic equation, which is infinitely stiff. For that reason, it is recommended to solve by an \emph{implicit} method, e.g., the \emph{Crank-Nicolson} method ($\theta=1/2$) or the \emph{backward-Euler} method ($\theta=1$). The former has a convergence of order two, whereas the latter has a convergence of order one, meaning that the time integration error asymptotically decreases by a factor four or two, respectively, when dividing the time step in two. On the other hand, the backward-Euler method has nicer stability properties which result from numerical damping of the solution. The backward-Euler method is used in the accompanying exercise \cite{De-Gersem_2019ac,De-Gersem_2019ad}. However, also more sophisticated time integrators, as provided in, e.g., \cite{Treichl_2015aa}, can be exploited.

It has been shown that magnetoquasistatic FE simulation can benefit from more sophisticated time-integration methods.
\begin{itemize}
  \item \emph{Higher-order} time integrators, e.g., from the family of Runge-Kutta methods \cite{Hairer_1996aa}, can achieve a~much higher convergence order for smooth problems. Moreover, they allow to construct an~\emph{embedded} solution, i.e., a solution with a lower convergence order, which can be used for error estimation and thereby enables an error-controlled adaptive selection of the time step \cite{Clemens_2002ab}. While for the classical Euler method, one has to use simpler approaches, e.g., one can compare the solution for one time step of size $\Delta t_{k+1}$ with the result of two time steps of size $\Delta t_{k+1}/2$. If they differ much, one reduces the time steps, whereas, if they are very close, one may enlarge it \cite{Hairer_1996aa}.
  \item In some models, phenomena at two largely different time scales occur. Then, \emph{multirate} time-integration techniques can be employed which perform time-stepping for each phenomenon at its own rate \cite{Pels_2018aj}.
  \item In case of models consisting of several components, e.g., multi-physical simulations or methods coupling different discretization techniques, \emph{co-simulation} with \emph{waveform relaxation} can bring a~significant improvement \cite{Schops_2010aa}. The method iterates the solutions for several sub-problems obtained by independent solvers on a common time window until convergence. The convergence of the waveform iteration has to be proven by numerical analysis \cite{Miekkala_1987aa}.
  \item Although counter-intuitive, time integration can be done in parallel by so-called \emph{parallel-in-time} or \emph{parareal} methods \cite{Maday_2003aa,Gander_2007aa}. The method minimizes the discrepancies occurring at the time instants between the consecutive time windows when a time integrator is applied in parallel, by a~type of shooting method. To obtain an efficient algorithm, one need to dispose of a \emph{coarse} time integrator which may be less accurate but should be much faster than the \emph{fine} time integrator. Recent extensions dedicated to time-periodic problems have been proposed \cite{Gander_2013ab,Kulchytska-Ruchka_2019ac}.
\end{itemize}

\section{Ferromagnetic saturation, linearization of the formulation}
\label{sect:hdg_linearization}

The yoke parts of normal-conducting and superconducting magnets are made of steel or as a stack of steel laminates, the latter to prevent eddy currents along the direction perpendicular to the laminates. The~magnetic fields at which accelerator magnets are operated cause the steel to saturate. This nonlinearity has to be modelled and simulated accurately in order to get realistic values for the magnet's performance.

\subsection{Material model}

\begin{figure}[tb]
  \centering
  \includegraphics[width=11cm]{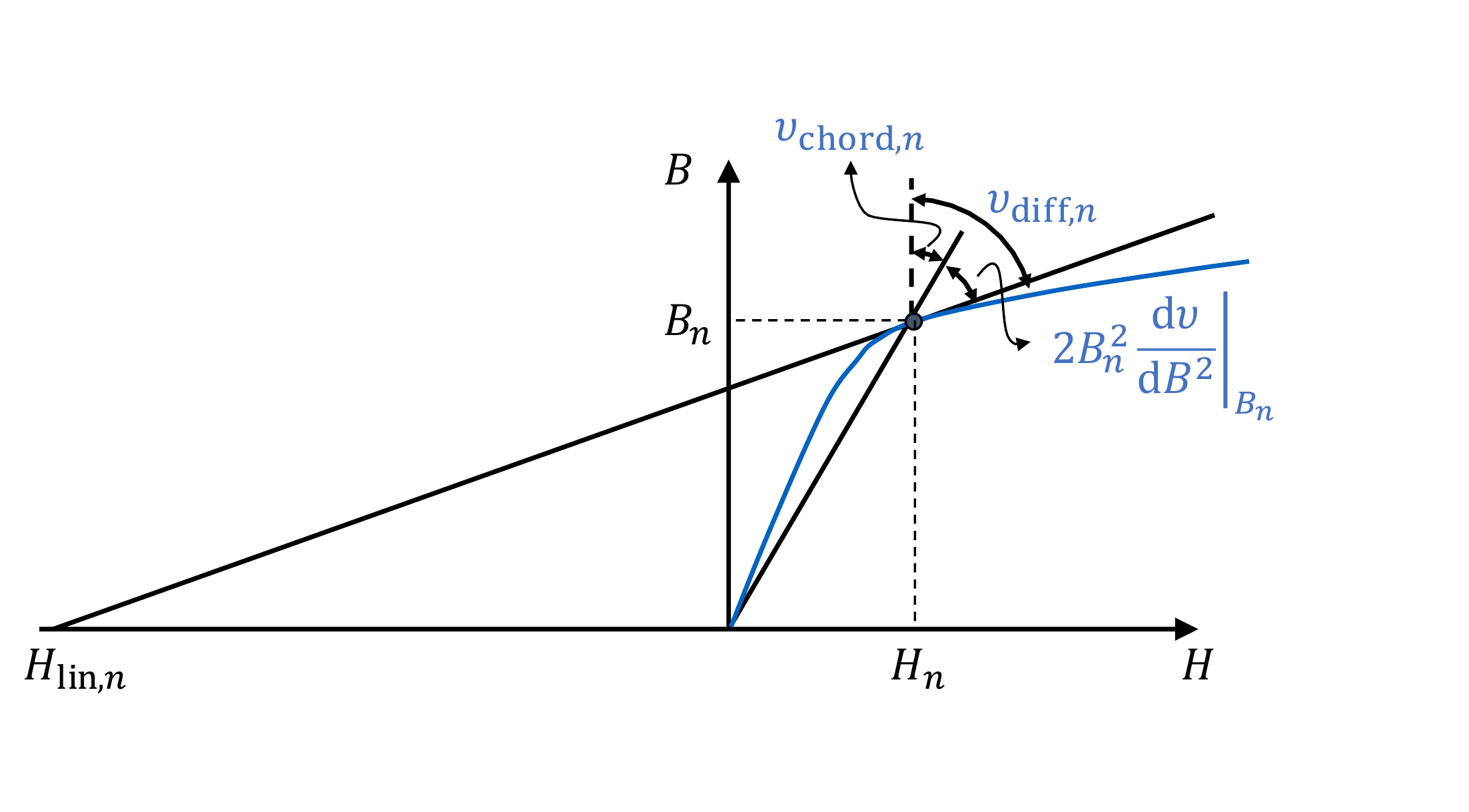}
  \caption{Operation point $(H_n,B_n)$ at a nonlinear BH-characteristic; chord reluctivity $\nu_{\text{chord},n}$ and the differential reluctivity $\nu_{\text{diff},n}$.}
  \label{fig:hdg_BH}
\end{figure}

In general, under time-varying operation, the material traverses outer and inner hysteresis loops. For many steel materials, however, it is acceptable to consider the \emph{anhysteretic} curve, i.e., the curve centred within the outer hysteresis loop, for the FE simulation itself, and to calculate the hysteresis losses according to specifications provided by the material vendors in a post-processing step. Here, an isotropic steel material is considered, represented by the expression $H=H(B)$. At each point in space and at each instant of time, the material is operated at an operation point $(H_n,B_n)$ on the anhysteretic curve (Fig.~\ref{fig:hdg_BH}). For the 1D case, one distinguishes between the \emph{chord reluctivity} $\nu_{\text{chord},n}=H_n/B_n$ and the \emph{differential reluctivity} $\nu_{\text{diff},n}=\frac{\text{d}H}{\text{d}B}\big|_{B_n}$, which are related to each other by
\begin{align}
  \nu_{\text{diff},n} &=\nu_{\text{chord},n}+2B_n^2\frac{\text{d}\nu}{\text{d}B^2}\big|_{B_n} \,,
\end{align}
where $\frac{\text{d}\nu}{\text{d}B^2}$ follows from the anhysterestic curve. In the 3D case, the chord reluctivity remains scalar, whereas the differential reluctivity becomes tensorial, i.e.,
\begin{align}
  \overline{\overline{\nu}}_{\text{diff},n} &=\nu_{\text{chord},n}\overline{\overline{1}} +2\vec{B}_n\frac{\text{d}\nu}{\text{d}B^2}\big|_{B_n}\vec{B}_n \,,
\end{align}
with $\overline{\overline{1}}$ the unit tensor. The tensorial form of the differential reluctivity illustrates a phenomenon called \emph{cross magnetization}, occurring under saturation, even for isotropic materials \cite{De-Gersem_2008aa}.

\subsection{Homogenization of a lamination stack}

Typically lamination stacks of accelerator magnets are several cm up to several m in length and consist of lamination sheets with a thickness in the order of $1$~mm, featuring a coating at both sides of $10$-$100$~$\mu$m. As a consequence, it is unmanageable to resolve the individual lamination sheets in the overall 3D model. Instead, the lamination stack is modelled by a bulk part with homogenized material parameters. Here, we assume the stacking in the $z$-direction. The ratio of steel with respect to the full volume is characterized by a \emph{stacking factor} $\gamma_\text{pk}$.

The eddy-current effect in the lamination stack is modelled by the conductivity tensor
\begin{align}
\overline{\overline{\sigma}} &=\text{diag}\left(\gamma_\text{pk}\sigma,\gamma_\text{pk}\sigma,0\right) \,,
\end{align}
where $\sigma$ is the conductivity of the steel material. Because of the nonlinearity of steel, the homogenization of the magnetic effects is more complicated. The laminations themselves are made from an isotropic steel with a characteristic given by the expressions $H_\text{Fe}=H_\text{Fe}(B_\text{Fe})$ or, equivalently, $B_\text{Fe}=B_\text{Fe}(H_\text{Fe})$. A~magnetic flux along the $x$- and $y$-direction experiences a parallel connection of steel and non-permeable coating, whereas a magnetic flux along the $z$-direction traverses a series connection thereof. This behaviour is represented by two adapted $BH$-characteristics:
\begin{subequations}
\begin{align}
B_{xy}(H_{xy}) &=\gamma_\text{pk}B_\text{Fe}(H_{xy}) +(1-\gamma_\text{pk})\mu_0 H_{xy} \,;\\
H_z(B_z)       &=\gamma_\text{pk}H_\text{Fe}(H_z)    +(1-\gamma_\text{pk})\nu_0 B_z  \,.
\end{align}
\end{subequations}

\subsection{Linearization}

The nonlinearity forces to repeat the solution of system of equations for updated values for the reluctivities. Many techniques for solving the system of nonlinear equations exist \cite{Nocedal_2006aa,Ortega_2000aa}. Here, the two most common techniques, i.e., the successive-substitution method and the Newton method are described.

The \emph{successive-substitution} approach uses the linearized material relation
\begin{align}
  \textcolor{red}{\vec{H}} &=\textcolor{blue}{\nu_{\text{chord},n}}\textcolor{red}{\vec{B}} \,,
\end{align}
whereas the \emph{Newton} approach uses the linearized material relation (Fig.~\ref{fig:hdg_BH})
\begin{align}
  \textcolor{red}{\vec{H}} &=\textcolor{blue}{\vec{H}_{\text{lin},n}} +\textcolor{blue}{\overline{\overline{\nu}}_{\text{diff},n}}\textcolor{red}{\vec{B}} \,.
\end{align}

When inserted in the magnetostatic formulation (insertion in the magnetodynamic formulation is similar), one gets
\begin{align}
  \nabla\times\left(\textcolor{blue}{\nu_{\text{chord},n}}\nabla\times\textcolor{red}{\vec{A}_{n+1}^*}\right) &=\vec{J} \,;\\
  \label{eq:hdg_newton}\nabla\times\left(\textcolor{blue}{\overline{\overline{\nu}}_{\text{diff},n}}\nabla\times\textcolor{red}{\vec{A}_{n+1}^*}\right) &=\vec{J}-\nabla\times\textcolor{blue}{\vec{H}_{\text{lin},n}} \,,
\end{align}
for the successive-substitution and Newton approaches, respectively. The Newton method formulated as in Eq. \eqref{eq:hdg_newton} is readily implemented when the FE assembly procedures allow to consider tensorial reluctivities and arbitrary magnetization terms \cite{Koch_2008ad}.

Because the convergence of a naive version of successive substitution is poor, a new iterate for the magnetic vector potential is typically obtained by relaxation, i.e.,
\begin{align}
  \textcolor{red}{\vec{A}_{n+1}} &=\alpha\textcolor{red}{\vec{A}_{n+1}^*} +(1-\alpha)\textcolor{blue}{\vec{A}_n} \,,
\end{align}
with relaxation factor $\alpha<1$ \cite{Pechstein_2004aa}. For Newton's method, similar techniques are known, e.g., line-search methods and trust-region methods  \cite{Nocedal_2006aa,Vande-Sande_2003aa}. However, in practice, Newton tends to work even without relaxation for many practical problems. For both methods, the convergence of the nonlinear iteration is monitored by checking a relevant criterion. E.g., the nonlinear iteration is stopped when the change in magnetic energy between two successive nonlinear iteration steps drops below a user-defined tolerance.

\section{System solution}
\label{sect:hdg_system_solution}

The magnetostatic formulation as well as the magnetodynamic formulation with implicit time stepping leads after linearization to a large but sparse algebraic system of equations. The system is symmetric and positive (semi-)definite. Up to several millions of DoFs, the most reliable and fast solution method is a sparse direct solver, which is essentially based on the well-known idea of Gaussian elimination \cite{Schenk_2004aa}. For larger systems, iterative solvers, such as, e.g., the preconditioned conjugate gradients (CG) method, are needed \cite{Saad_2000aa}. As a preconditioner, an (algebraic) multigrid technique is recommended \cite{Mertens_1998aa,Trottenberg_2001aa,Reitzinger_2002aa,Bochev_2003aa}. On parallel computing systems, domain-decomposition methods \cite{Smith_1996ab} together with a load-balanced partitioning of the mesh \cite{Karypis_1998aa} is used. For repeated solutions, an improved convergence of the iterative solver may be achieved by exploiting deflation techniques \cite{De-Gersem_2001ad,Clemens_2004aa}.

\section{Post-processing}
\label{sect:hdg_postprocessing}
Several quantities of interest (QoIs) of accelerator magnets can be derived from the solution for the~magnetic vector potential.

\subsection{Magnetic flux density}
\label{subsect:hdg_magnfluxdens}

The magnetic flux density $\vec{B}=\nabla\times\vec{A}$ is calculated element-wise from the magnetic vector potential $\vec{A}_h$ (see also in the exercise \cite{De-Gersem_2019ac,De-Gersem_2019ad}). Because the FE method employs polynomial shape functions, this operation decreases the approximation order by~$1$. This is fully acceptable for visualization purposes but may be inacceptable when the field values are needed themselves. A way out is to apply local post-processing techniques avoiding or repairing for the loss of accuracy, e.g., by \emph{defect correction} \cite{Romer_2017ad}.

\subsection{Current density}
\label{subsect:hdg_currdens}

The current density is given by 
\begin{align}
  \label{eq:hdg_current} \vec{J} &=\sum_{q=1}^{n_\text{coil}} \vec{\chi}_q i_q +\vec{J}_\text{eddy} \,,
\end{align}
where $\vec{J}_\text{eddy}=-\sigma\frac{\partial\vec{A}}{\partial t}$ is the eddy-current density. Here, the derivative with respect to time also leads to a~loss of accuracy, which can be compensated by a higher-order time integrator or by correction techniques.

\subsection{Magnetic energy}

The magnetic energy $W_\text{magn}$ follows by integrating the magnetic energy density $w_\text{magn}$ at the computational mesh. For nonlinear steel, the result of $w_\text{magn}=\int_0^{\vec{B}}\vec{H}\cdot\text{d}\vec{B}$ follows from the material curve, whereas for the linear parts, $w_\text{magn}=\frac{1}{2}\nu B^2$ and for linear permanent-magnet material, one can use $w_\text{magn}=\frac{1}{2}\nu(B_r-B)^2$, which agrees with all existing definitions of the magnetic energy density in a~magnetized material up to an arbitrary constant. The magnetic energy is primarily stored in the air and vacuum parts experiencing large magnetic fields. For a magnet without permanent magnets and with a~single coil carrying an instantaneous current $I$, the coil's (chord) inductance follows from
\begin{align}
  L_\text{chord} &=\frac{2W_\text{magn}}{I^2} \,.
\end{align}
\pagebreak
\subsection{Joule loss}

The Joule loss consists of several contributions:
\begin{enumerate}
  \item The \emph{ohmic loss in the coils} of the magnet is
  \begin{align}
    P_\text{Ohm} &=\sum_{q=1}^{n_\text{coil}} R_q i_q^2 \,.
  \end{align}
  \item The \emph{eddy-current loss} $P_\text{eddy}$ \emph{in the conducting parts} can be integrated from the eddy-current loss density $p_\text{eddy}=\frac{1}{\sigma}J_\text{eddy}^2$. Alternatively, it can be found directly from
  \begin{align}
         P_\text{eddy} &=\frac{\text{d}u^T}{\text{d}t}\mathbf{M}_\sigma\frac{\text{d}u}{\text{d}t} \,.
  \end{align}
  \item The \emph{eddy-current loss in the lamination stack} can be calculated in a post-processing step or inserted in the model by the approach described in \cite{Gyselinck_1999aa}.
  \item The \emph{hysteresis loss} $P_\text{hyst}=\int_\Omega p_\text{hyst}\;\text{d}V$ can be estimated using the Steinmetz-Bertotti formula for the hysteresis loss density
  \begin{align}
    p_\text{hyst} &=k_\text{hyst}\frac{f}{50\;\text{Hz}} \left(\frac{|\vec{B}|^2}{1\;\text{T}}\right)^2 \,,
  \end{align}
  where $k_\text{hyst}$ is a constant related to the particular material and $f$ is the main frequency of operation.
\end{enumerate}
Especially for the dimensioning of the cryostat and the cooling system of a superconducting magnet, an accurate calculation of the Joule loss, probably beyond the simple methods sketched above, may be necessary.

\subsection{Aperture field quality}

For accelerator magnets, the quality of the magnetic field distribution in the aperture is of paramount importance. For dipole magnets used for deflecting the particle beam, the magnetic field should be as homogeneous as possible. On the other hand, for quadrupole magnets used for focusing the beam, the magnetic field obviously should be close to a pure quadrupole field. Because the evaluation of local magnetic flux densities comes together with a loss of accuracy (see Section~\ref{subsect:hdg_magnfluxdens}), a dedicated post-processing tool is applied for characterizing the aperture field. This approach is here explained for the~2D case.

From the FE solution, the $z$-component of the magnetic vector potential is evaluated at a circle with reference radius $r_0$ lying in the aperture and centred around the beam axis. The data are represented  by the Fourier coefficients $a_p$ and $b_p$:
\begin{align}
  A_z(r_0,\varphi) &=\sum_{p=0}^\infty \left(a_p\cos(p\varphi) +b_p\sin(p\varphi)\right) \,.
\end{align}
The magnetic vector potential in the aperture is then characterized by \cite{Russenschuck_2010aa}
\begin{align}
  A_z(r,\varphi) &=\sum_{p=0}^\infty \left(a_p\cos(p\varphi) +b_p\sin(p\varphi)\right) \left(\frac{r}{r_0}\right)^p \,.
\end{align}
The magnetix flux density is
\begin{subequations}
\begin{align}
  B_r(r,\varphi) &=\sum_{p=1}^\infty \frac{p}{r}\left(-a_p\sin(p\varphi) +b_p\cos(p\varphi)\right) \left(\frac{r}{r_0}\right)^p \,;\\
  B_\varphi(r,\varphi) &=\sum_{p=1}^\infty \frac{p}{r} \left(-a_p\cos(p\varphi) -b_p\sin(p\varphi)\right) \left(\frac{r}{r_0}\right)^p \,.
\end{align}
\end{subequations}
When evaluated at $r=r_0$, one finds for the radial component of the magnetic flux density
\begin{align}
  B_r(r_0,\varphi) &=\sum_{p=1}^\infty \frac{p}{r_0} \left(-a_p\sin(p\varphi) +b_p\cos(p\varphi)\right) \,;\\
  B_r(r_0,\varphi) &=\sum_{p=1}^\infty \left(B_p\sin(p\varphi) +A_p\cos(p\varphi)\right) \,,
\end{align}
where $B_p$ and $A_p$ are called the \emph{normal} and \emph{skew} \emph{multipole coefficients} given in tesla at reference radius $r_0$, which can be calculated directly from the Fourier coefficients $a_p$ and $b_p$. In some solvers, the calculation of the multipole coefficients is tightly integrated in the field solver itself \cite{De-Gersem_2002ac} (see also Section~\ref{subsect:hdg_fese}).

The quality of the aperture field of a $2P$-pole magnet (for a dipole magnet, $P=1$) measured at the reference radius is given by
\begin{align}
  Q &=\sum_{p=1,p\neq P}^\infty \frac{B_p^2+A_p^2}{B_P^2} \,,
\end{align}
where $B_P$ denotes the normal multipole coefficient of the nominal magnetic field.

\subsection{Fringe field and stray fields}

\emph{Fringe} fields correspond to the broadening of the magnetic flux path when traversing the aperture between the poles, whereas \emph{stray} fields are fields swarming around the device, thereby possibly disturbing nearby equipment. Fringe fields are counteracted by an appropriate design of the magnetic path and, in particular, the pole shoes. Stray fields are reduced by \emph{passive} (highly permeable or highly conducting shields) and/or \emph{active} (current-carrying coils) \emph{magnetic shielding}. Field simulation allows the evaluation of fringe and stray fields. Stray fields are typically quantified with respect to the main magnet field and expressed in a logarithmic scale.

\subsection{Computational quench detection}

In superconducting magnets, the superconducting wires inavoidably experience the own magnetic field and the magnetic field generated by nearby coils. Because quench is initiated when a certain threshold is locally exceeded, field simulation is used to evaluate the probability of quench due to too high DC and AC magnetic fields in the wire regions. One can go a step further and also perform thermal field calculations, from which one also can judge the probability of quench due to local hot spots. By that, simulating accelerator magnets becomes a highly complicated multiscale and multiphysics simulation task. The development of appropriate simulation techniques is still a matter of ongoing research \cite{Bortot_2018aa} (see also Section~\ref{subsect:hdg_quench}).

\section{Modelling and simulating an accelerator magnet}
\label{sect:hdg_modelling}

The process of modelling an accelerator magnet is typically carried out with the help of a graphical user interface (GUI) for computer aided design (CAD) and computer aided engineering (CAE). The geometry is defined in a tool for solid modelling or imported from such a tool. Sometimes, existing CAD data contain many details which are only relevant for construction purposes or for mechanical and thermal simulations. It may be worthwhile to discard such details in order to simplify the electromagnetic model and reduce the computation time. It is also recommended to parametrize the model as far as needed. This allows semi-automatic parameter studies and optimization steps to be carried out later on.

In a second step, materials are defined or selected from a material database, and assigned to the different regions of the model. At the front and back magnet side, a complicated interplay is expected between the ferromagnetic-saturation and eddy-current effects in the lamination stacks. Hence, there, a~realistic modelling of the materials and composite materials is necessary. 
As excitations, the currents applied to the coils are specified.

A tedious task is the definition of boundary conditions (BCs) to be applied at the model boundaries. For iron-dominated magnets, one assumes that all magnetic flux is contained with the iron hull. Then, electric BCs, which corresponds to Dirichlet BCs when using a magnetic-vector-potential formulation, are correct. For coil-dominated magnets, however, considerable stray fluxes may occur. Then, one should lay the model boundary at a sufficient distance, use so-called \emph{open BCs} or apply a combination of both strategies. Many magnets feature particular mirror symmetries, which can be exploited to reduce the~model size by a factor two, four or eight. Symmetry planes at which the magnetic flux lines are oriented tangentially, are modelled by electric BCs, whereas symmetry planes which are traversed by magnetic flux lines perpendicularly, are modelled by magnetic BCs, which correspond to homogeneous Neumann BCs for the magnetic-vector-potential formulation.

A crucial step is the construction of the computational mesh. Although many sofware packages provide a fully automated mesher, possibly combined with adaptive mesh refinement routines called during the solution stage, it is recommended to do a few simulation tests on beforehand, e.g., for a~single time instant, thereby monitoring the convergence of the most important QoIs according to the size of the mesh. An accurate and at the same time affordable simulation may be obtained after having manually specified the mesh density in some crucial parts of the model, especially where eddy currents are expected (the mesh must resolve the skin depth with several element layers).

\section{Example: SIS-100 dipole magnet}
\label{sect:hdg_example}

\subsection{Context}

\begin{figure}[tb]
  \centering
  \includegraphics[width=6.5cm]{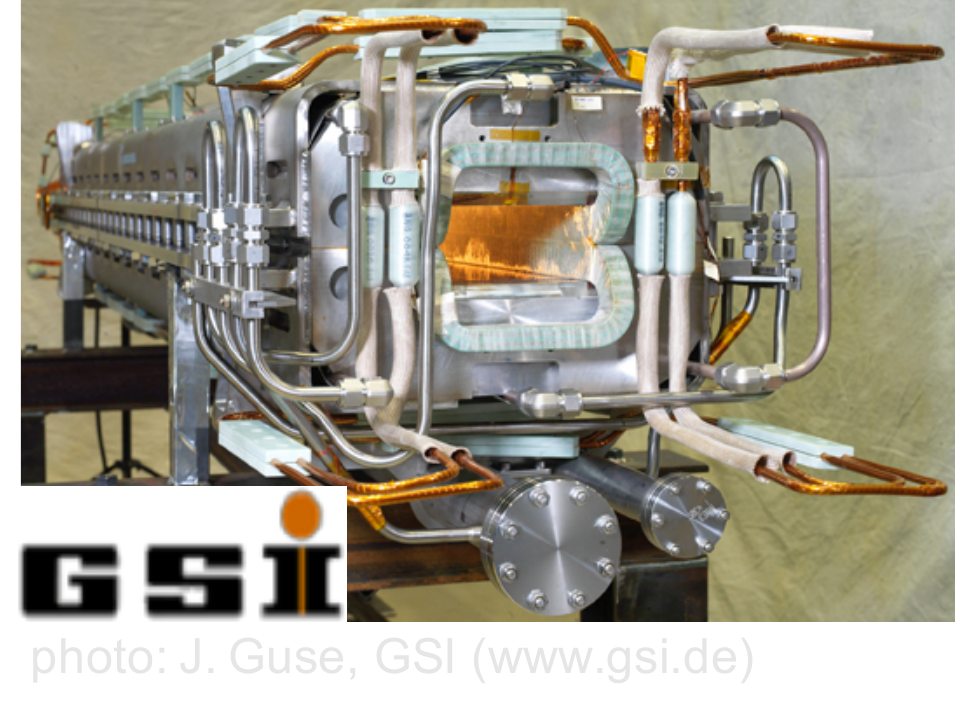}
  \hspace{1cm}\includegraphics[width=7.5cm]{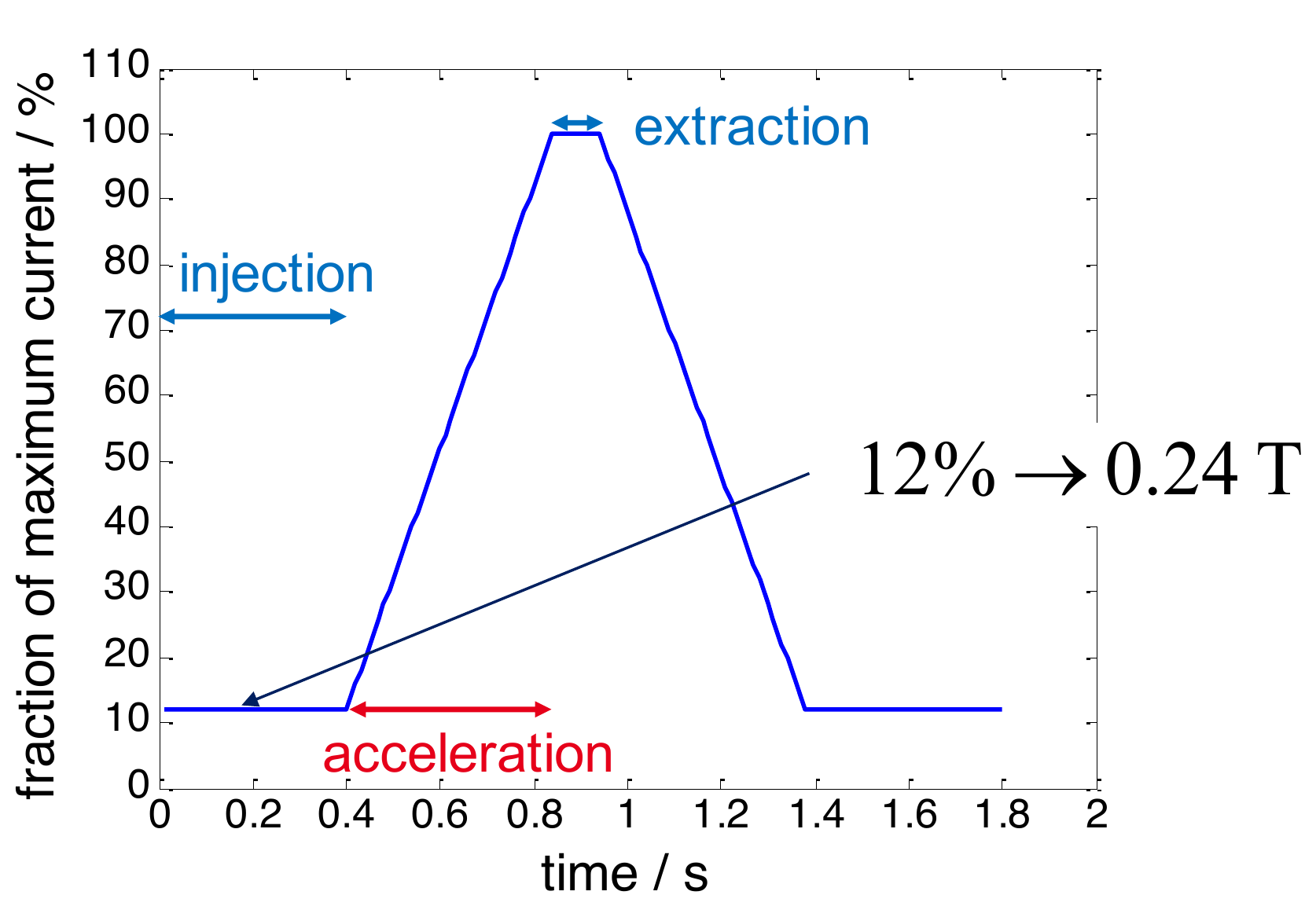}
  \caption{Excitation cycle of the SIS-100 magnet.}
  \label{fig:sis100_cycle}
\end{figure}

In this section, a 3D nonlinear transient simulation of a superconducting dipole magnet is given as illustration. The emphasis is here put on computational aspects. The simulation has been carried out for determining the eddy-current loss per cycle in the SIS-100 magnet \cite{Koch_2006aa,Koch_2008aa,Koch_2008ad,Koch_2009ab}, which has been designed as the main dipole for the SIS-100 synchrotron of the Facility for Antiproton and Ion Research (FAIR) \cite{FAIR_2016aa}, which is currently under construction at the Helmholtzzentrum für Schwerionenforschung (GSI, Facility for Heavy Ion Research) \cite{GSI_2016aa} in Darmstadt, Germany. The SIS-100 magnet has a nominal aperture dipole field of $2$~T. The magnet is ramped at a rate of $4$~T/s (Fig.~\ref{fig:sis100_cycle}), which comes together with significant eddy-current and hysteresis losses in the ferromagnetic yoke. The quantification and minimization of these losses during design was of primordial importance for minimizing the magnet's operation cost and for dimensioning the cryostat. The yoke is laminated in order to prevent eddy currents which would otherwise compensate for the time-changing currents in the coils. As a consequence, the~main eddy-current effect will occur at the front and back magnet sides due to leakage flux leaving the~lamination stack perpendicularly. The calculation of this effect necessitates a 3D FE model.

\subsection{Spatial discretization}

\begin{figure}[tb]
  \centering
  (a)\includegraphics[height=3.0cm]{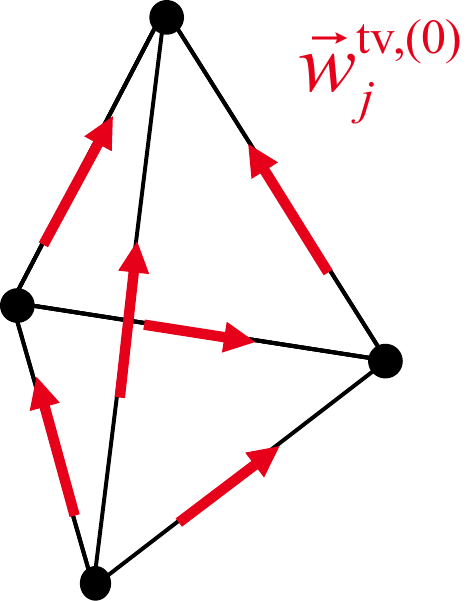}
  (b)\includegraphics[height=3.0cm]{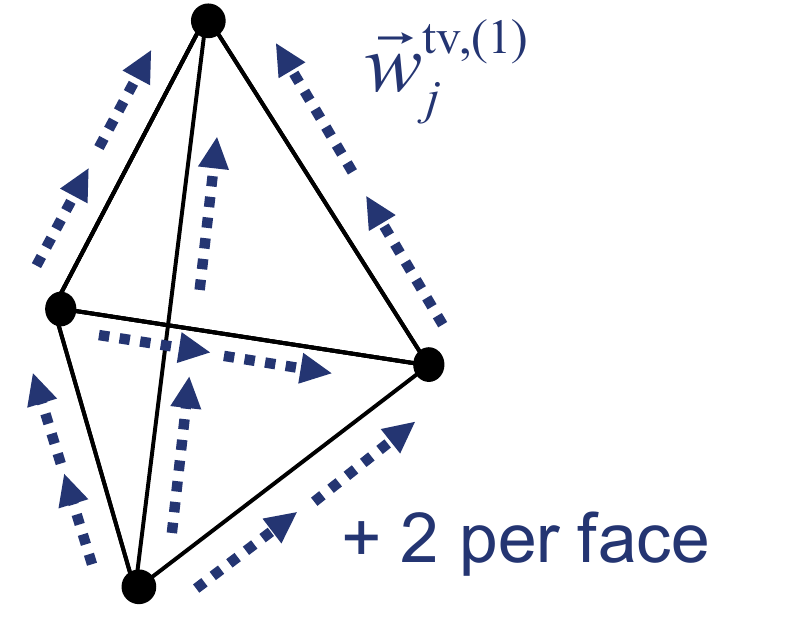}
  \caption{Allocation of the degrees of freedom for (a) first-order and (b) second-order edge elements.}
  \label{fig:sis100_tets}
\end{figure}

The model uses a 3D tetrahedral mesh with a user-defined mesh density (Fig.~\ref{fig:sis100_mesh}b). The mesh is constructed by the meshing routines embedded in CST DESIGN SUITE \cite{CST_2017aa}. A 3D nonlinear transient FE solver has been built on top of the FEMSTER library and TRILINOS algebraic tool set \cite{Castillo_2005aa,Heroux_2005aa,Koch_2008aa}. Both first-order edge elements ($6$ DoFs associated with the edges of a tetrahedron) and second-order edge elements ($12$ DoFs associated with the edges and $8$ DoFs associated with the faces of a tetrahedron) are used to discretize the magnetic vector potential (Fig.~\ref{fig:sis100_tets}). The time integration is carried out by the backward-Euler method \cite{Koch_2008ad}. The nonlinear problem is linearized by the Newton method \cite{Koch_2008aa}. 

\subsection{Solution}

\begin{figure}[tb]
  \centering
  (a)\includegraphics[width=6.4cm]{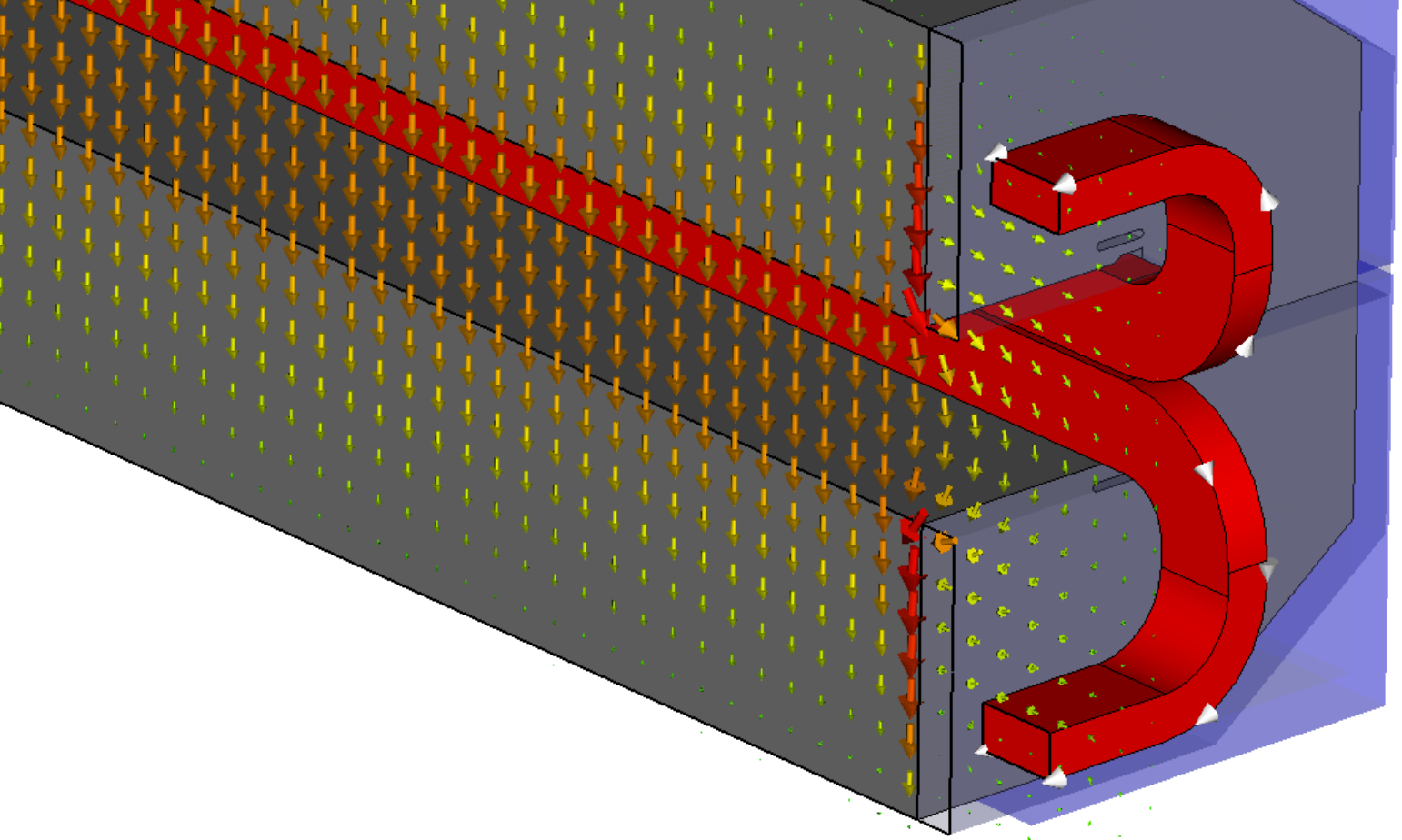}\hspace{1cm}
  (b)\includegraphics[width=6.4cm]{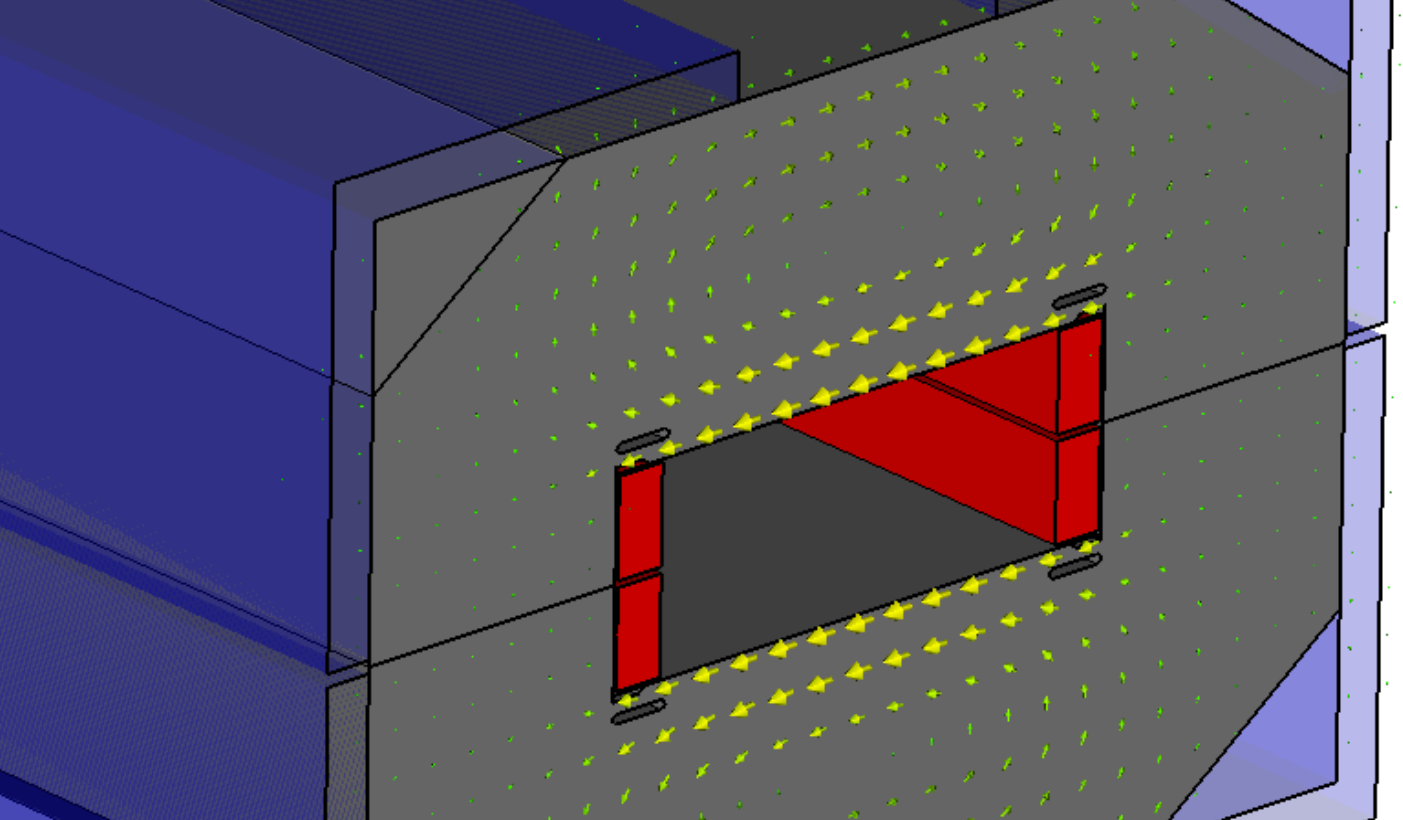}
  \caption{(a) Magnetic flux in the aperture and (b) eddy currents at the front of the SIS-100 magnet.}
  \label{fig:sis100_fields}
\end{figure}

\begin{figure}[tb]
  \centering
  \includegraphics[width=7.0cm]{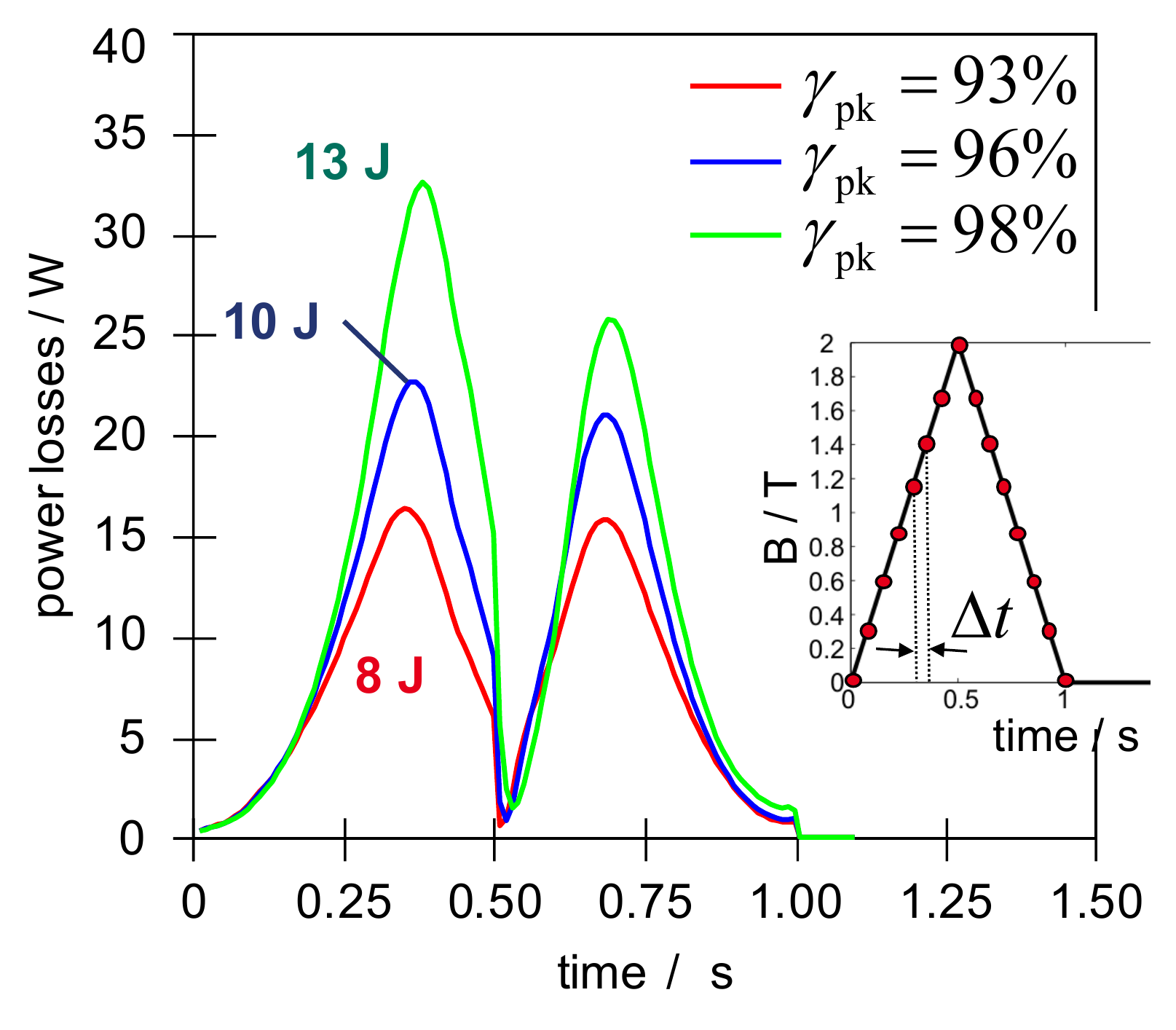}
  \caption{Eddy-current loss in the ferromagnetic yoke of the SIS-100 magnet as a function of the stacking factor $\gamma_\text{pk}$.}
  \label{fig:sis100_loss}
\end{figure}

The magnetic flux in a midplane of the magnet and the eddy currents in the first lamination of the yoke
are shown in Fig.~\ref{fig:sis100_fields}. The space-integrated eddy-current loss in the magnet yoke has been simulated for different stacking factors (Fig.~\ref{fig:sis100_loss}). The time-integrated loss energies amount to $8$~J, $10$~J and $13$~J for a~stacking factor of $93\%$, $96\%$ and $98\%$, respectively. One clearly observes the down-ramp and up-ramp time span. The asymmetry is related to the magnetization time constant. One should keep in mind that these losses occur in a cold yoke at $4$~K and thus require considerable effort to be cooled away.

\subsection{Convergence of the spatial discretization error}

\begin{figure}[tb]
  \centering
  (a)\includegraphics[width=7.0cm]{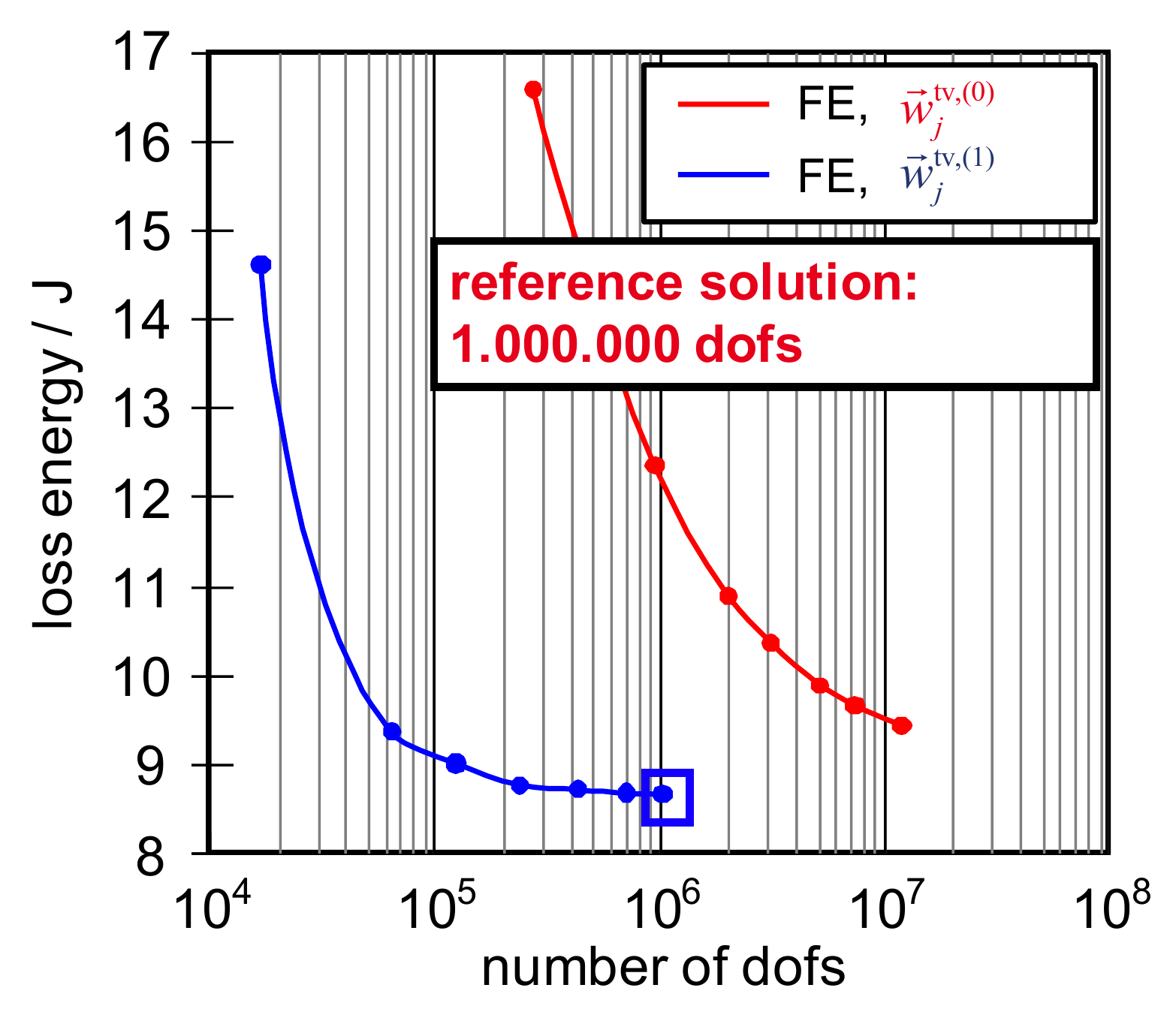}
  (b)\includegraphics[width=7.5cm]{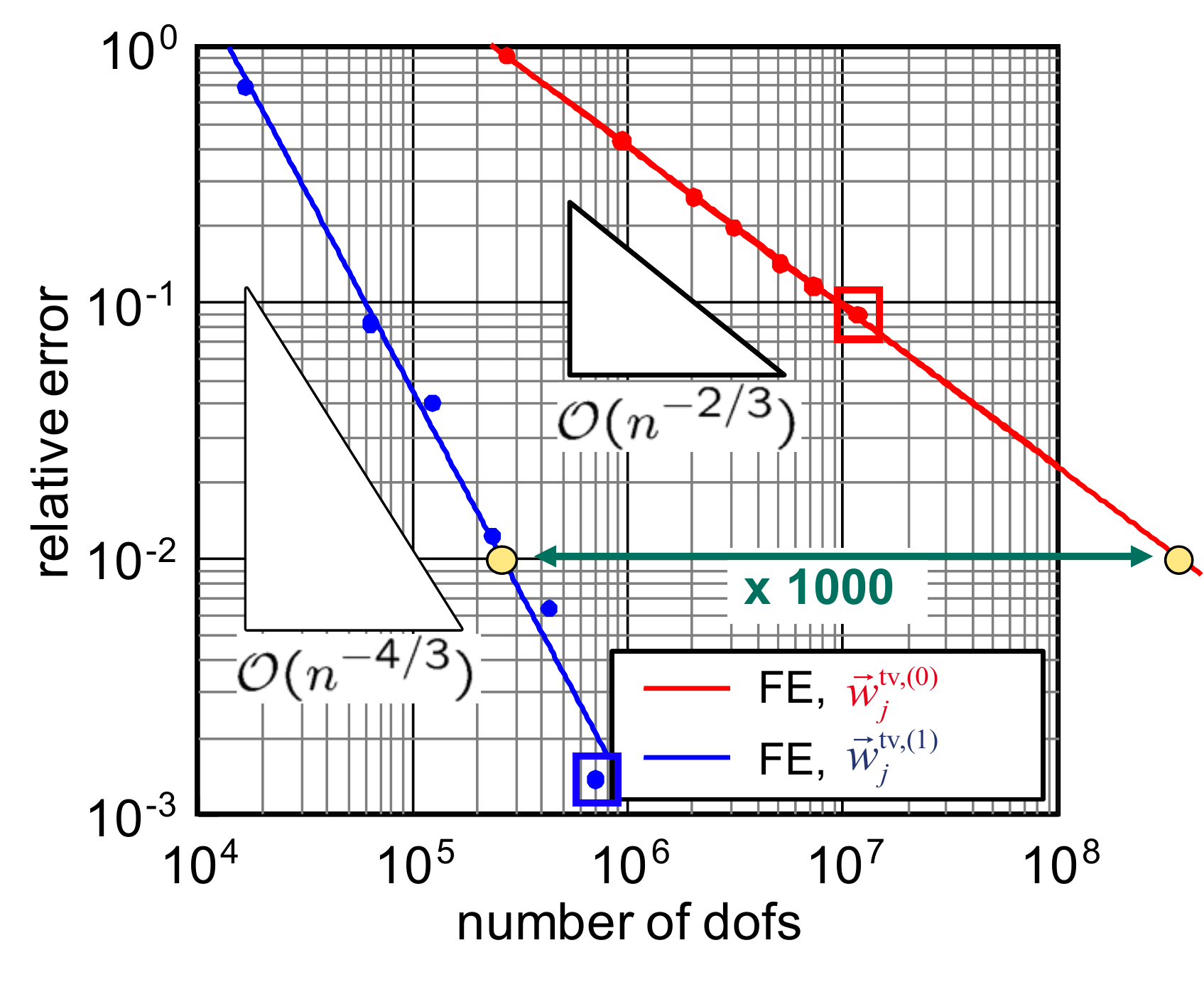}
  \caption{Convergence of the spatial discretization error measured for the Joule loss in the lamination stack of the SIS-100 magnet: (a) convergence towards the Joule loss; (b) relative error; the red square corresponds to a transient FE simulation with first-order edge elements on $132$~CPUs taking $4.4$~h of simulation time, whereas the blue square corresponds to a transient FE simulation with second-order edge elements on $72$~CPUs taking $7.3$~h of simulation time. The calculations have been carried out on a distributed computing system with 200 nodes, 400 CPUs, clock speed $2.4$~GHz, 2400 cores and 3200 GB of total memory.}
  \label{fig:sis100_convergence}
\end{figure}

A convergence study for the spatial discretization error has been carried out (Fig.~\ref{fig:sis100_convergence}). As expected, the~solver with second-order edge elements converges faster than the one with first-order edge elements. To obtain an accuracy of $1\%$ for the Joule loss, the second-order method needs almost $300000$ DoFs, which is a factor $1000$ less than the first-order method (Fig.~\ref{fig:sis100_convergence}a). In fact, only the second-order method is capable of attaining an accuracy of $1\%$ with an affordable amount of DoFs. However, this picture changes a bit when the computation time is taken as the decisive criterion for comparison. The calculations are carried out on a distributed computing system with $200$~nodes, $400$~CPUs, clock speed $2.4$~GHz, $2400$~cores and $3200$~GB of total memory. An optimal number of CPUs is selected by hand. The overall 3D nonlinear transient FE simulation takes $4.4$~hours on $132$~CPUs for first-order edge elements (red square in Fig.~\ref{fig:sis100_convergence}(b), attaining an accuracy of only $10\%$ or $7.3$~h on $72$~CPUs for second-order edge elements, almost reaching an accuracy of $0.1\%$ (blue square in Fig.~\ref{fig:sis100_convergence}(b)). This numerical test shows that higher-order edge elements should be preferred, but also that further parallelization may not bring further benefits. Third-order edge elements are even less parallelizable and are therefore not efficient.

3D nonlinear transient magnetoquasistatic field simulutions of accelerator magnets can be carried out in all commercial software packages mentioned in the introduction. The performance of these packages increases with the years, e.g., by the increasing parallelization of all parts of the simulation procedure. Typically, one needs to get familiar with the particular way of operating the specific tool and one needs to customize the post-processing routines in order to obtain all QoIs which are relevant for accelerator magnets.

\section{Advanced modelling and simulation techniques for accelerator magnets}
\label{sect:hdg_advanced}

The ongoing quest for more accurate and faster 3D transient field simulation of accelerator magnets triggers further improvements. Here, a few examples of recent developments are reported on.

\subsection{Field-circuit coupling}

When the magnet coils are excited by prescribed voltages rather than prescribed currents, or even more generally, if the behaviour of the excitation circuit needs to be simulated in close relation to the field model, a \emph{field-circuit coupled} formulation is required. The magnetoquasistatic formulation Eq. \eqref{eq:hdg_Astarform} is accompanied by a set of circuit equations. If the coil voltages $u_q(t)$ are prescribed, the additional equations read
\begin{align}
  R_q i_q +\int_\Omega\frac{\partial\vec{A}}{\partial t}\cdot\vec{\chi}_q\;\text{d}V &= u_q \,,\qquad q=1,\ldots,n_\text{coil} \,,
\end{align}
where $R_q$ are the DC resistances of the coils \cite{Schops_2013aa}. If a more general external circuit is considered, a~coupling between the field equations and a circuit modelled by modified nodal analysis (MNA) is set up \cite{Cortes-Garcia_2019aa}.

\subsection{Improved modelling of the aperture}
\label{subsect:hdg_fese}

The aperture field needs to be simulated with a high precision in order to predict harmonic distortion factors which are expected to be in the range of $10^{-4}$. Besides a-posteriori accuracy improvement techniques such as, e.g., defect correction \cite{Romer_2017ad}, there exists the possibility to a-priori select a high-precision discretization technique for the aperture region. While for the yoke parts, the FE method is more or less inavoidable because of the material nonlinearity, the overall method becomes hybrid, which may necessitate the development of a dedicated algebraic solution technique to retain the simulation efficiency \cite{De-Gersem_2003ab}.

\begin{figure}
  \centering
  (a)\includegraphics[width=5.8cm]{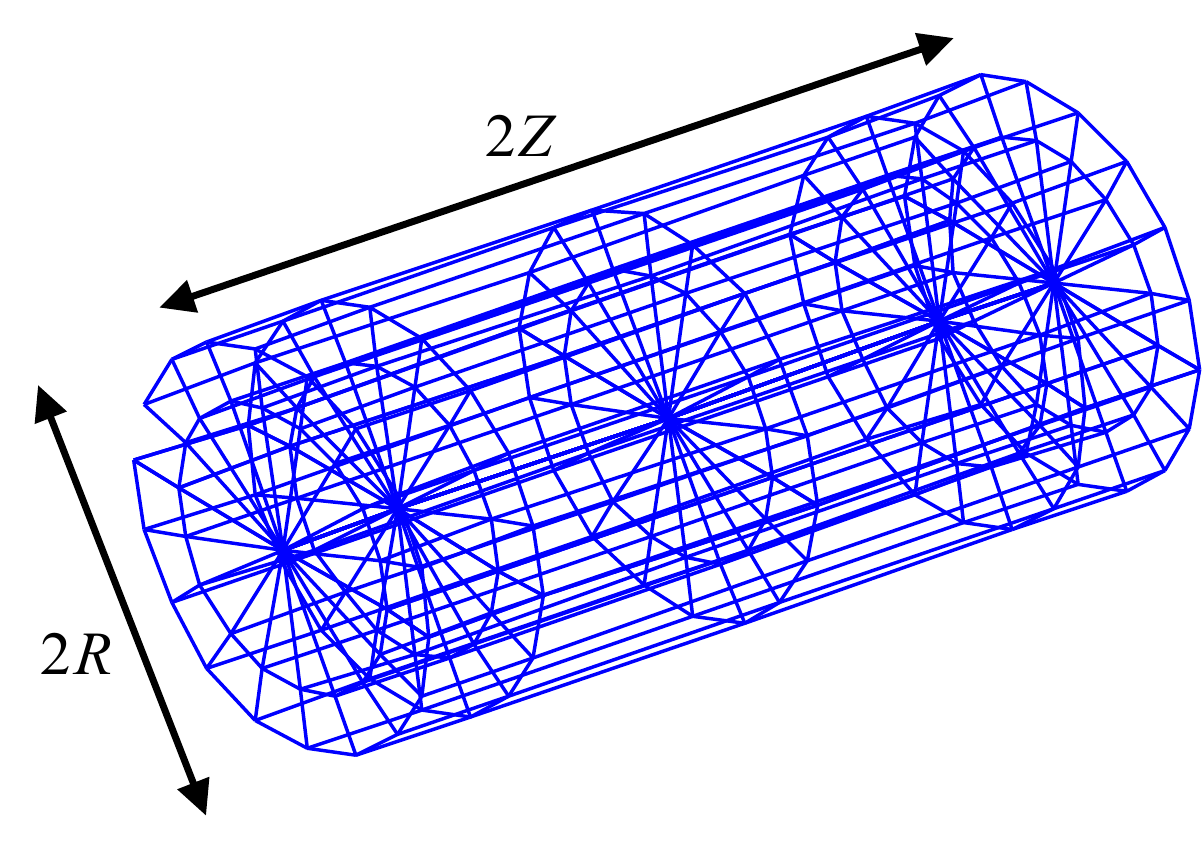}
  (b)\includegraphics[width=4.2cm]{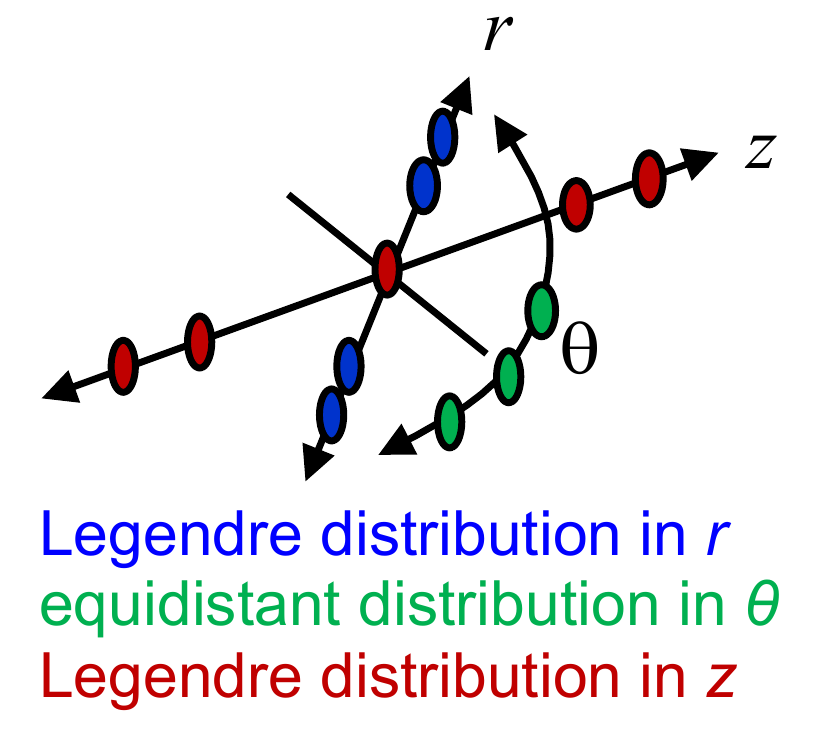}
  (c)\includegraphics[width=3.8cm]{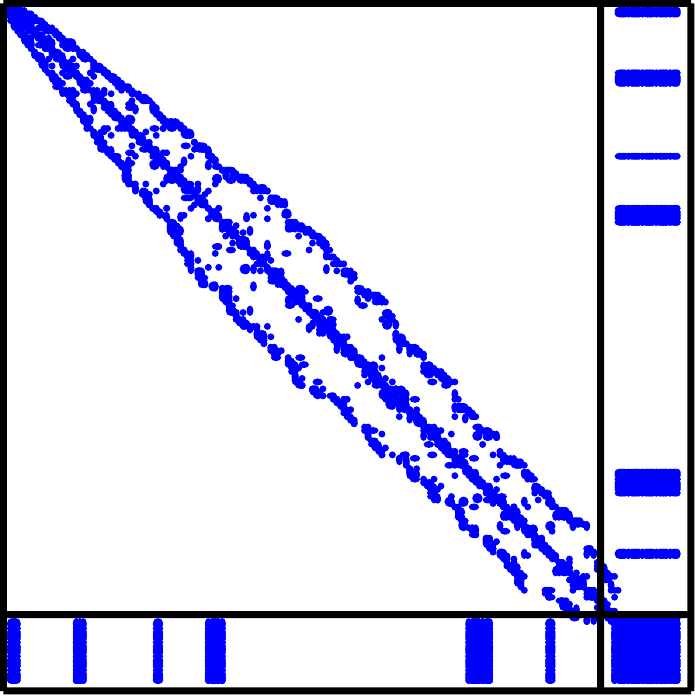}
  \caption{FE-SE hybrid discretization technique: (a) SE grid in the aperture region; (b) distribution of the SE collocation nodes; (c) hybrid system of algebraic equations.}
  \label{fig:fese}
\end{figure}

In ROXIE, a FE-boundary-element coupling is used \cite{Kurz_2000aa}. Another possibility is to insert a spectral-element (SE) discretization in the cylindrical aperture \cite{Dehler_1994aa,De-Gersem_2002ac}. The idea of the method is as follows. The~magnetic field in the aperture is modelled by the magnetic scalar potential, which is discretized by means of SE shape functions of the form
\begin{align}
  N_{m,p,n}(r,\varphi,z) &=P_m\left(\frac{r}{R}\right) e^{\jmath p\varphi} P_n\left(\frac{z}{Z}\right) \,,
\end{align}
where $P_m$ denotes the Legendre polynomial of degree $m$, $R$ is the radius and $2Z$ is the length of the~aperture region (Fig.~\ref{fig:fese}). At the interface between the FE region and the SE region, appropriate interface conditions are formulated. The resulting system of algebraic equations consists of a large sparse part corresponding to the FE model part and a small fully populated part corresponding to the SE model part (Fig.~\ref{fig:fese}(c)).

\subsection{Beam pipe}

\begin{figure}
  \centering
  \includegraphics[width=16cm]{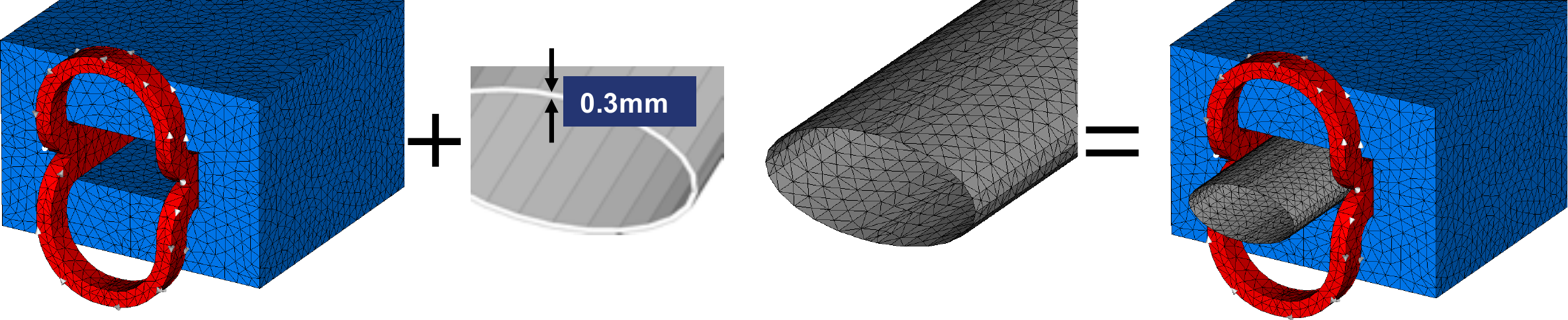}
  \caption{Simplified SIS-100 model with a beam pipe modelled by shell elements associated with a triangulation of the surface of the beam pipe.}
  \label{fig:sis100_beampipe}
\end{figure}

The beam pipe within an accelerator magnet is conducting and thus carries eddy currents during the~ramping of the magnet. Resolving a thin beam pipe by the 3D mesh may not be the most efficient approach. Instead, shell elements, which only requires the surface of the beam pipe to be resolved by the mesh, are applied \cite{Poignard_2008aa,Koch_2009ad} (Fig.~\ref{fig:sis100_beampipe}).

\begin{figure}
  \centering
  \includegraphics[width=5.5cm]{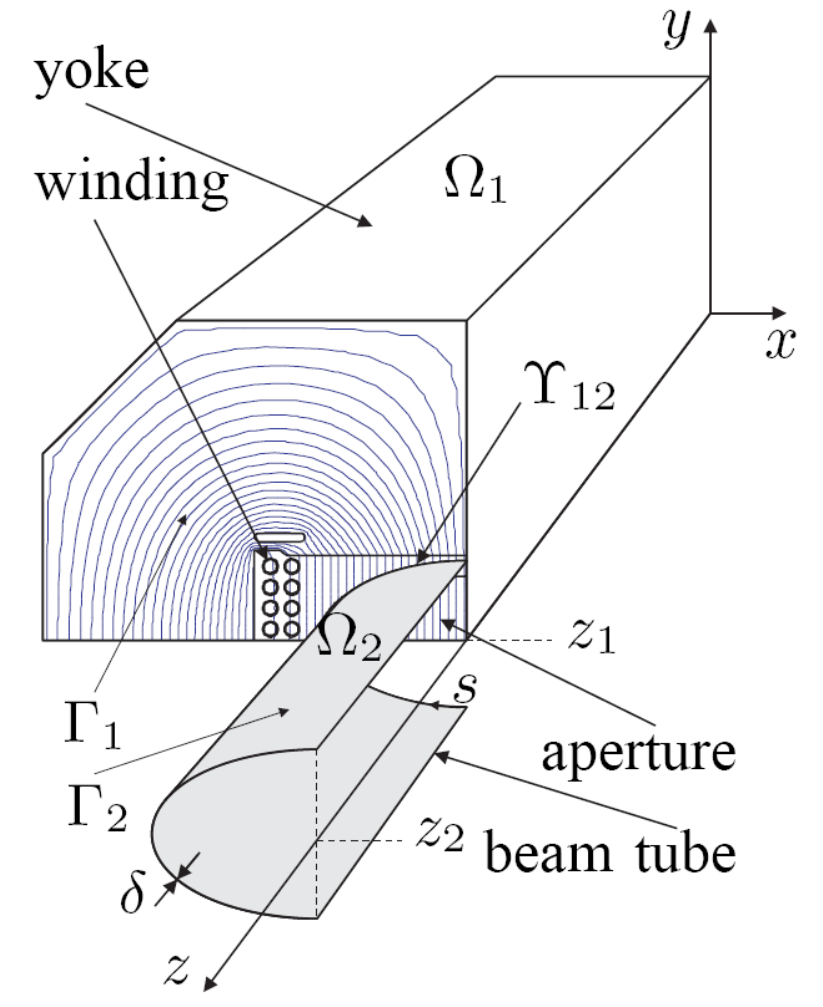}
  \caption{2D FE model of the SIS-100 magnet coupled to a stationary-current formulation modelling the closing paths of the eddy currents induced in the beam pipe.}
  \label{fig:sis100_beltrami}
\end{figure}

In 2D magnet models, one should take care of the closing paths of the eddy currents at the front and back sides of the magnet, which are responsible for a considerable fraction of the resistance of the~current path. To that purpose, formulations exist which couple the 2D magnetoquasistatic model of the magnet's cross section to a stationary-current model discretized on the beam-pipe surface, modelling the closing paths (Fig.~\ref{fig:sis100_beltrami}) \cite{De-Gersem_2009ac}.

\subsection{Uncertainty quantification}

Many parameters of an accelerator-magnet model suffer from statistical variations. Material parameters may be exposed to processing variability, whereas geometric parameters may vary due to construction tolerances. Well-established methods can be used to propagate these uncertainties from the input parameters to statistical measures (mean value, variance) for the QoIs. The Monte-Carlo method is generally applicable but typically requires more FE model evaluations than affordable. Methods using generalized polynomial chaos (gPC), e.g., stochastic collocation \cite{Babuska_2007aa} and stochastic Galerkin \cite{Ghanem_2003aa} methods, allow to carry out uncertainty quantification (UQ) at reasonable computational expenses \cite{Bartel_2013ab,Romer_2014aa}.

\subsection{Geometric optimization}

Further improvement of magnet systems requires the optimization of their geometries, especially the~yoke shapes and the shapes and positions of the coils \cite{Hemker_2008aa}. This requires a parametrization of the geometry and a high spatial resolution of the FE simulation train. The former can be achieved by design elements or by mapping techniques \cite{Ion_2018aa}, whereas the latter is achieved by increasingly fine meshes, or by applying improved discretization techniques such as, e.g., isogeometric analysis (IGA) \cite{Pels_2015aa}.

\subsection{Laminated yoke parts}

So far, the laminated yoke parts are considered as a bulk model part. The insulation between the laminates is modelled by putting the conductivity in the direction perpendicular to the laminates to zero and by scaling the $BH$-curve for the ferromagnetic material down by the stacking factor. The complicated interaction between the impinging field, ferromagnetic saturation and the eddy-current effect is thereby not accurately modelled. Further improvement can be achieved by better homogenization techniques \cite{Dular_2003aa,De-Gersem_2012aa,Hollaus_2018aa} or, in general, by multiscale models \cite{Kevrekidis_2009aa,Niyonzima_2018aa}.

\subsection{Ferromagnetic-material modelling}

The behaviour of ferromagnetic material in high magnetic fields, under large mechanical stresses and for a large temperature range is very complicated. It is very difficult to span the full range by a few curves based on measurement results, especially when the measurement data do not span the full operation range. As a way out, there is a tendency to employ micro-magnetic models in combination with multiscale techniques \cite{Vanoost_2016aa}. This combination is still challenging but will become a realistic option within nearby future when better computational homogenization techniques and more computational resources become available.

A further difficulty is caused by the degradation brought to ferromagnetic material during the~shaping process, e.g., by punching \cite{Bali_2016aa} or by laser cutting \cite{Bali_2017aa}. The elastic deformation or thermal transitions come together with a significant deterioration of the magnetic properties, which needs to be taken into account, especially for smaller magnet systems where the fraction of damaged material is comparably large.

\subsection{Modelling AC losses in superconducting cable}

The presence of an AC magnetic field causes additional losses in superconducting cable. One distinguishes between persistent currents in the superconducting filaments, inter-filament coupling currents and cable eddy currents \cite{Wilson_1987aa}. Resolving these effects on a 3D mesh of the full magnet system is impossible. Alternatively, homogenized cable models are used \cite{Verweij_1993aa,Takacs_1995aa,De-Gersem_2004ae}.

\subsection{Quench simulation}
\label{subsect:hdg_quench}

Superconducting magnets need to be protected against quench \cite{Wilson_1987aa}. When an initiating quench is detected, additional heat is inserted, either by heaters or by additional losses caused by an intentionally added current excitation. The aim is to induce quench in an as large as possible region, such that the~complete magnet quenches and provides a sufficiently large resistive voltage drop to throttle the current. Both the detection of a local quench and the procedure of quenching the full magnet need to be simulated accurately. The necessary components of such a simulation tool include a magnetoquasistatic field solver, a thermal field solver, an accurate model for the superconducting material and a master algorithm for coupling and time integration. A current research project dedicated to this meta-task is the ''Simulation of Transient Effects in Accelerator Magnets'' (STEAM) project \cite{Bortot_2018ab,STEAM_2019aa}. The main idea of STEAM is to rely upon well-established tools and to achieve a coupling between the components by efficient mesh interpolation \cite{Maciejewski_2018aa} and by well-designed waveform-relaxation approaches \cite{Cortes-Garcia_2017ab}.
 
\section{Summary and conclusions}

This lecture note explains the basics of the finite-element (FE) method as applied for the magnetic field simulation of accelerator magnets. The mathematics behind the FE method are derived in its most simple form. Details with high relevance for magnet simulation are pointed out. It is possible to implement a~simple 2D FE solver for magnet simulation within a few hours, as shown in the accompanying exercise. Nevertheless, attaining a sufficient accuracy within an affordable simulation time for the general 3D nonlinear and transient case requires a customized and highly optimized solver, as illustrated for the example of the SIS-100 magnet. The note furthermore gives an overview of contemporary efforts for further improving accelerator-magnet simulation.

\section*{Acknowledgements}

We wish to thank Wolfgang Ackermann, Bernhard Auchmann, Thorben Casper, Erion Gjonaj, Stephan Koch, Iryna Kulchytska-Ruchka, Dimitrios Loukrezis, Nicolas Marsic, Andreas Pels, Stephan Russenschuck, Jens Trommler and Arjan Verweij for their contributions in this field of research.

\end{document}